%% file: gw230529_lensing.tex
\documentclass[usenatbib, twocolumn]{mnras}
\usepackage{calc}
\usepackage{mathtools,graphicx}
\usepackage{pifont}
\usepackage{bm}
\usepackage{microtype}
\usepackage{booktabs}
\usepackage{times}
\usepackage[varg]{txfonts}
\usepackage[utf8]{inputenc}
\usepackage[caption=false]{subfig}
\usepackage{paralist}
\usepackage[normalem]{ulem}
\usepackage{multirow}
\usepackage{etoolbox}
\usepackage{longtable}
\usepackage{xstring}
\usepackage{xparse}
\usepackage[left]{lineno}
\usepackage{blindtext}
\usepackage[nolist,nohyperlinks,printonlyused]{acronym}
\usepackage{dcolumn}
\usepackage{diagbox}
\usepackage{mathrsfs}
\usepackage{amsmath,amsfonts,amssymb,pifont,gensymb}
\usepackage{multirow,enumerate}
\usepackage{natbib}
\usepackage{slashbox}
\usepackage{xcolor}

\definecolor{NOTECOLOR}{rgb}{0.4, 0.2, 0.1}
\definecolor{notecolor}{rgb}{0.4, 0.2, 0.1}
\definecolor{darkgreen}{rgb}{0.1, 0.8, 0.2}
\definecolor{maroon}{HTML}{AF3235}

\newcommand{\boldtheta}{\boldsymbol{\theta}}
\newcommand{\boldphi}{\boldsymbol{\phi}}
\newcommand{\LambdaTilde}{\tilde{\Lambda}}

\newcommand{\dcc}{P2400353}

\title[Lensing hypothesis for GW230529]{What is the nature of GW230529? An exploration of the gravitational lensing hypothesis}
\input{author_list.tex}

\date{\today -- LIGO-\dcc}
\pubyear{2024}

\begin{document}

\maketitle

\clearpage

\begin{abstract}
    \input{abstract.tex}
\end{abstract}


\section{Introduction}\label{sec:intro}
\input{introduction.tex}

\section{A Brief Description of GW230529}\label{sec:gw230529}
\input{description_gw230529.tex}

\section{Gravitational Lensing of Gravitational Waves}\label{sec:lensing}
\input{lensing.tex}

\section{Rates and Properties of Lensed BNS Events}\label{sec:rates}
\input{rates.tex}

\section{Population Considerations for a Lensed GW230529}\label{sec:population}
\input{population.tex}

\section{Properties of GW230529 Under the Lensing Hypothesis and Consistency with Observations}\label{sec:delensing_effect}
\input{delensing.tex}

\section{Extra Checks for Lensing Features}\label{sec:extra_checks}
\input{extra_checks.tex}

\section{Conclusions}\label{sec:conclusions}
\input{conclusions.tex}

\section*{Acknowledgements}
\input{acknowledgements.tex}

\bibliographystyle{aasjournal}
\bibliography{bibliography.bib}

\end{document}

%% file: author_list.tex
\author[Janquart, Keitel and Lo et al.]{
Justin Janquart$^{1,2,3,4}$\thanks{justin.janquart@uclouvain.be},
David Keitel$^{5,6}$\thanks{david.keitel@ligo.org},
Rico K.~L.~Lo$^{7}$\thanks{kalok.lo@nbi.ku.dk},
Juno C.~L.~Chan$^{7}$,
Jose Mar\'ia Ezquiaga$^{7}$, \newauthor
Otto A. Hannuksela$^{8}$,
Alvin K.~Y.~Li$^{9,8,10}$,
Anupreeta More$^{11,12}$,
Hemantakumar Phurailatpam$^{8}$,\newauthor
Neha Singh$^{5}$,
Laura E.~Uronen$^{8}$,
Mick Wright$^{13}$,
Naresh Adhikari$^{14}$,
Sylvia Biscoveanu$^{15}$,
Tomasz Bulik$^{16}$, \newauthor
Amanda M.~Farah$^{17}$, 
Anna Heffernan$^{5}$,
Prathamesh Joshi$^{18,19}$,
Vincent Juste$^{20}$,
Atul Kedia$^{21}$, \newauthor
Shania A.~Nichols$^{22}$, 
Geraint Pratten$^{23}$,
C.~Rawcliffe$^{24}$,
Soumen~Roy$^{4,3,1,2}$,
Elise M.~S\"anger$^{25}$,
Hui Tong$^{26,27}$, \newauthor
M.~Trevor$^{28}$,
Luka Vujeva$^{7}$,
Michael Zevin$^{29,15}$
\\
$^{1}$Center for Cosmology, Particle Physics and Phenomenology - CP3, Universit\'e Catholique de Louvain, Louvain-La-Neuve, B-1348, Belgium\\
$^{2}$Royal Observatory of Belgium, Avenue Circulaire, 3, 1180 Uccle, Belgium\\
$^{3}$Department of Physics, Utrecht University, Princetonplein 1, 3584 CC Utrecht, The Netherlands\\
$^{4}$Nikhef, Science Park 105, 1098 XG Amsterdam, The Netherlands\\
$^{5}$Departament de F\'isica, Universitat de les Illes Balears, IAC3–IEEC, E-07122 Palma, Spain\\
$^{6}$University of Portsmouth, Institute of Cosmology and Gravitation, Portsmouth PO1 3FX, United Kingdom\\
$^{7}$Niels Bohr International Academy, Niels Bohr Institute, Blegdamsvej 17, DK-2100 Copenhagen, Denmark\\
$^{8}$Department of Physics, The Chinese University of Hong Kong, Shatin, New Territories, Hong Kong \\
$^{9}$LIGO Laboratory, California Institute of Technology, Pasadena, CA 91125, USA\\
$^{10}$RESCEU, The University of Tokyo, Tokyo, 113-0033, Japan \\
$^{11}$The Inter-University Centre for Astronomy and Astrophysics (IUCAA), Post Bag 4, Ganeshkhind, Pune 411007, India \\
$^{12}$Kavli Institute for the Physics and Mathematics of the Universe (IPMU), 5-1-5
Kashiwanoha, Kashiwa-shi, Chiba 277-8583, Japan \\
$^{13}$SUPA, School of Physics and Astronomy, University of Glasgow, Glasgow, Scotland \\
$^{14}$Leonard E. Parker Center for Gravitation, Cosmology, and Astrophysics, University of Wisconsin–Milwaukee, Milwaukee, WI 53201, USA \\
$^{15}$Center for Interdisciplinary Exploration and Research in Astrophysics (CIERA), Northwestern University, 2145 Sheridan Road, Evanston, IL, 60201, USA \\
$^{16}$Astronomical Observatory, University of Warsaw, Aleje Ujazdowskie 4, 00478 Warsaw,  Poland\\
$^{17}$Department of Physics, University of Chicago, Chicago, IL 60637, USA\\
$^{18}$Department of Physics, The Pennsylvania State University, University Park, PA 16802, USA\\
$^{19}$Institute for Gravitation and the Cosmos, The Pennsylvania State University, University Park, PA 16802, USA\\
$^{20}$Service de Physique Th\'eorique, CP225, Universit\'e Libre de Bruxelles, Boulevard du Triomphe, 1050 Bruxelles, Belgium\\
$^{21}$Center for Computational Relativity and Gravitation, Rochester Institute of Technology, Rochester, New York 14623, USA\\
$^{22}$Louisiana State University, Baton Rouge, LA 70803, USA\\
$^{23}$School of Physics and Astronomy, University of Birmingham, Edgbaston, Birmingham, B15 2TT, United Kingdom\\
$^{24}$University of British Columbia, Vancouver, BC V6T 1Z4, Canada\\
$^{25}$Max Planck Institute for Gravitational Physics (Albert Einstein Institute), Am M{\"u}hlenberg 1, Potsdam, 14476, Germany\\
$^{26}$School of Physics and Astronomy, Monash University, VIC 3800, Australia\\
$^{27}$OzGrav: The ARC Centre of Excellence for Gravitational Wave Discovery, Clayton VIC 3800, Australia\\
$^{28}$University of Maryland, College Park, Maryland 20742, USA\\
$^{29}$The Adler Planetarium, 1300 South DuSable Lake Shore Drive, Chicago, 60605, IL, USA\\
}

%% file: abstract.tex
On the 29$^{\mathrm{th}}$ of May 2023, the LIGO--Virgo--KAGRA Collaboration
observed a compact binary coalescence event
consistent with a neutron star -- black hole merger,
though the heavier object of mass $2.5-4.5\,M_{\odot}$
would fall into the purported lower mass gap.
An alternative explanation for apparent observations of events in this mass range has been suggested as
strongly gravitationally lensed binary neutron stars.
In this scenario, magnification would lead to
the source appearing closer and heavier than it really is.
Here, we investigate the chances and
possible consequences for the GW230529 event to be gravitationally lensed.
We find this would require high magnifications
and we obtain low rates for observing such an event,
with a relative fraction of lensed versus unlensed observed events of $2\times 10^{-3}$ at most.
When comparing the lensed and
unlensed hypotheses accounting for the latest rates and population model, we find a $1/58$ 
chance of lensing, disfavoring this option.
Moreover, when the magnification is assumed to be strong enough
to bring the mass of the heavier binary component below
the standard limits on neutron star masses,
we find high probability for the lighter object to have a sub-solar mass,
making the binary even more exotic than a mass-gap neutron star--black hole system.
Even when the secondary is not sub-solar, its tidal deformability would likely be measurable,
which is not the case for GW230529.
Finally, we do not find evidence for extra lensing signatures such
as the arrival of additional lensed images, type-II image dephasing, or microlensing.
Therefore, we conclude it is unlikely for 
GW230529 to be a strongly gravitationally lensed binary neutron star signal.

%% file: introduction.tex
GW230529\_181500~\citep[for short GW230529,][]{LIGOScientific:2024elc}
is a gravitational-wave (GW) signal from a compact binary coalescence (CBC)
observed by the LIGO--Virgo--KAGRA collaboration (LVK)
close to the start of its fourth observing run (O4a).
Its component masses made GW230529 unusual,
different from any events in the previous 
GWTC releases~\citep{LIGOScientific:2021djp}
from the Advanced LIGO~\citep{LIGOScientific:2014pky},
Advanced Virgo~\citep{VIRGO:2014yos} 
and KAGRA~\citep{Somiya:2011np, Aso:2013eba, Akutsu:2020his} network.
The standard interpretation is for the signal
to originate from the merger of a neutron star with an object 
between $2.5$ and $4.5\,M_{\odot}$~\citep{LIGOScientific:2024elc}. 
While confirming the exact nature of the heavier object is difficult
with GW data alone, it seems most likely the source is not a 
binary neutron star (BNS) but rather a
neutron star--black hole (NSBH) merger, with a light black hole in the 
so-called lower mass gap~\citep{LIGOScientific:2024elc}. 
Observations of low-mass X-ray binaries in the Galaxy do not show evidence
of stellar mass black holes in this region of the mass spectrum
\citep{Ozel:2010kle, Farr:2011bde},
and it is difficult to populate with standard stellar progenitor models.
However, there is now growing evidence
that this mass region is not completely empty~\citep{KAGRA:2021duu,LIGOScientific:2024elc}.

An alternative explanation for GW events observed with component masses 
in the lower mass gap is
gravitational lensing of regular BNSs~\citep{Bianconi:2022etr,Magare:2023hgs,Canevarolo:2024muf}. 
For magnified events, the inferred luminosity distance
is lower than the true one, leading to an underestimation of the source
redshift and hence overestimation of the source-frame masses. 
Therefore, the heavier component of a regular BNS could be mistaken for 
a mass-gap compact object.
Without an electromagnetic counterpart, it is generally difficult to confirm the nature of the 
event, with or without lensing~\citep{Bianconi:2022etr}. 
An exception occurs  
when the tidal deformation of the binary leaves an imprint on the GW waveform,
since the lensing hypothesis can be verified when the measured deformability is incompatible 
with the apparent masses for a realistic equation of state (EoS)~\citep{Pang:2020kle}.
Other options include finding multiple lensed copies (images) of GWs from the same source~\citep[e.g.,][]{Haris:2018vmn},
overlapping signals in the high-magnification limit~\citep{Lo:2024wqm},
and frequency-dependent lensing effects on the waveform~\citep{Takahashi:2003ix},
their image type~\citep{Dai:2017huk,Ezquiaga:2020gdt}, or finding additional microlensing features~\citep[e.g.,][]{Mishra:2021xzz}.
Due to the limited current BNS detection range,
lensing rates for such events are however low,
as the lensing optical depth is small at low redshifts.  
This is why GW lensing searches have so far mainly focused on binary black holes (BBHs),
which can be detected from further away~\citep{Hannuksela:2019kle,LIGOScientific:2021izm, LIGOScientific:2023bwz,Janquart:2023mvf}

In this work, we investigate the possibility of GW230529 being a 
lensed BNS appearing as a NSBH merger when analysed under the unlensed hypothesis.
This work is structured as follows.
In Section~\ref{sec:gw230529}, we describe GW230529 in more
detail, including some discussions on the implications of the event 
if it is not lensed. In Section~\ref{sec:lensing}, we describe 
the basics of gravitational lensing of GWs, and explain how it 
can lead to biased estimates of the source parameters. Then, in 
Section~\ref{sec:rates}, we discuss the detection rates of 
lensed BNS signals with O4a sensitivity and
the expected characteristics of such events.
We then consider in Section~\ref{sec:population}
whether a lensed GW230529 would be more consistent with population models,
and compute a Bayes factor for this.
In Section~\ref{sec:delensing_effect}, we check explicitly what
the source-frame parameters of the event could be if it was lensed,
finding the source likely would be some other type of exotic binary system, 
including a sub-solar mass object.
Here, we also look at
whether we would expect to be able to measure
the tidal deformabilities for lensed signals matching GW230529.
In Section~\ref{sec:extra_checks}, 
we do extra checks to search for lensing features
including multiple images, negative image parity, and microlensing effects.
We summarize our findings and discuss the implications in Section~\ref{sec:conclusions}.

%% file: description_gw230529.tex
GW230529~\citep{LIGOScientific:2024elc} was observed on the 29$^\mathrm{th}$ of May 2023, 
at 18:15:00.7 UTC, by the LIGO Livingston detector. LIGO Hanford was
out of observing mode and Virgo was undergoing upgrades. KAGRA was in 
observing mode, but its sensitivity was too low to impact the analysis
of the event. The event was observed by three matched-filter pipelines:
\textsc{GSTLaL}~\citep{Messick:2016aqy,Sachdev:2019vvd,Hanna:2019ezx, Cannon:2020qnf,ewing2023performance,Tsukada:2023edh}, 
\textsc{PyCBC}~\citep{Allen:2004gu,Allen:2005fk,DalCanton:2020vpm,Usman:2015kfa,Nitz:2017svb,Davies:2020tsx}, and 
\textsc{MBTA}~\citep{Adams:2015ulm,Aubin:2020goo}, 
with similar signal-to-noise ratio (SNR) 
of about 11.4 in all of them.

Bayesian parameter estimation
under standard LVK priors\footnote{Using a BBH waveform with low-mass priors, and combining the results 
of different waveforms to mitigate potential systematics~\citep{LIGOScientific:2024elc}.}
gave 90\% credible intervals
of $201^{+102}_{-96}\,\mathrm{Mpc}$ for the luminosity distance $D_L$
and source-frame component masses of $3.6^{+0.8}_{-1.2}\,M_{\odot}$ and 
$1.4^{+0.6}_{-0.2}\,M_{\odot}$ for the heavier (primary) and lighter (secondary) component,
respectively. Therefore, the primary is likely in the mass-gap region,
taken as 3--5\,$M_{\odot}$~\citep{LIGOScientific:2024elc,Ozel:2010kle,Farr:2011bde},
and it would be of high interest to know whether it is a neutron star 
or a black hole.

Finding an electromagnetic counterpart could help answer this question.
However, as a single-detector observation, sky localization was poor,
making electromagnetic follow-ups difficult.
\citet{Chandra:2024ila} have reported that a BNS kilonova from the event
could have been detectable with a telescope like the future Rubin Observatory~\citep{LSST:2008ijt},
but no counterparts have been reported from telescopes available at the time,
with constraints reported in several wavelengths
\citep{Ronchini:2024lvb,Ahumada:2024qpr}.

The most direct GW-only test of the nature of the component objects
would be signatures of tidal deformability in the GW waveform.
The LVK analysis with a BNS waveform
~\citep[IMRPhenomPv2\_NRTidalv2,][]{Dietrich:2019kaq}
returned a posterior on the tidal deformability of the primary that peaks at zero,
compatible with a black hole.
Even though it falls off clearly towards higher values,
it is too broad to confidently exclude the possibility of a neutron star.
For the secondary, the posterior on the tidal deformability is uninformative, 
whether the event is analyzed with a BNS waveform
or one for NSBHs~\citep[IMRPhenomNSBH and SEOBNRv4\_ROM\_NRTidalv2\_NSBH,][]{Thompson:2020nei,Matas:2020wab}.
Therefore, one cannot tell whether the event 
is an NSBH or a BNS this way.

A more indirect approach is to check the spins of the components,
since neutron stars are expected to have spin amplitudes
below $0.05$~\citep{2008ApJ...672..479O}. However, spins 
are partially degenerate with masses~\citep{Cutler:1994ys}, and
difficult to measure individually for 
the two components~\citep{Vitale:2016avz, Shaik:2019dym}.
This made it challenging to tell whether the primary has a high spin or not.

Further tests were done to evaluate the probability of 
the primary being a neutron star under different hypotheses.
To do so, \citet{LIGOScientific:2024elc} looked at the probability 
of its mass being 
below the maximum allowed for a neutron star,
while marginalizing over the measured masses and spins, and different 
EoSs~\citep{KAGRA:2021duu, Essick:2020ghc}.
Allowing for large spins in the prior ($\chi_1, \chi_2 \leq 0.99$)\footnote{$\chi_i$ is 
the amplitude of the dimensionless spin vector of component $i$.},
a $2.9^{+0.4}_{-0.4}$\% probability was found for the primary to be a neutron star.
Restricting the spins to low values, i.e. $\chi_1, \chi_2 \leq 0.05$,
this probability dropped below 0.1\%.
Finally, accounting for a \textsc{Power Law + Dip + Break} population 
prior~\citep{Fishbach:2020ryj,Farah:2021qom, KAGRA:2021duu}, a 
probability of $8.8^{+2.8}_{-2.8}$\% was found,
taking into account that this prior leads to lower posterior mass estimates for the primary.
On the contrary, the probability of the secondary to be a neutron star
is higher than 95\% in all cases.
Similarly, \citet{Koehn:2024ape} obtained $\gtrsim84$\% probability
for the primary to be a black hole
using observationally constrained EoSs.
The impact of population-informed spin priors
was further investigated by \citet{Chattopadhyay:2024hsf},
finding consistent but potentially tighter mass estimates.
In summary, the preferred scenario under standard astrophysical assumptions
is that of an NSBH merger, 
with a black hole in the mass-gap region. 

Finally, assuming GW230529 came from an NSBH, there are also consequences 
for the mass distribution of black holes in such mergers,
with a decrease in the lowest expected mass of black holes, and an 
increased chance of having a black hole sub-population in the mass gap
\citep{LIGOScientific:2024elc}. 
This also  means that one needs to revisit the formation scenarios for such 
mergers to find compelling explanations for their observation.
Some such studies have already been conducted,
e.g. by \citet{Zhu:2024cvt,Chandra:2024ila,Ye:2024wqj}.
More exotic scenarios have also been considered, such as
primordial black holes~\citep{Huang:2024wse}.
The event has also already been used
to perform tests of general relativity \citep{Gao:2024rel,Sanger:2024axs,Julie:2024fwy},
where constraints are particularly tight
if the primary can be assumed to be a black hole.

%% file: lensing.tex
When a GW passes close to a massive object when travelling 
from the source to an observer, it can be gravitationally 
lensed~\citep{Ohanian1974, Degushi1986, Wang:1996as, Nakamura1998, Takahashi:2003ix},
modifying the observed waveform. The effect depends on the mass of the
deflector and the lens-source geometry. 
For the heaviest objects---galaxies or galaxy clusters---and 
best-aligned cases, we have strong 
lensing~\citep{Dai:2017huk,Haris:2018vmn, Ezquiaga:2020gdt,Liu:2020par,Lo:2021nae,Janquart:2021qov,Janquart:2023osz}.
In this case, the GW signal is split into multiple potentially detectable copies---referred to 
as ``images''---originating from the same sky location (within GW detector accuracy)
but magnified,
with a resolvable time delay, and with
an overall phase shift~\citep{Dai:2017huk,Ezquiaga:2020gdt}.
The regimes for
lower-mass lenses or less alignment
are often called millilensing,
where images with short time delays overlap in the detector band~\citep{Liu:2023ikc},
and microlensing,
where the waveform deformations due to lensing become frequency-dependent instead
\citep{Takahashi:2003ix,Wright:2021cbn}.

While the probability of detecting such phenomena is currently relatively low, 
it increases as detector sensitivity improves and as more detectors
are added to the global
network~\citep{Ng:2017yiu,Li:2018prc,Oguri:2018muv,Wierda:2021upe,Xu:2021bfn}.
Searches for lensing
signatures in the LVK data, focusing on BBH signals, 
have not yet provided
compelling evidence for 
lensing~\citep{Hannuksela:2019kle, Dai:2020tpj, LIGOScientific:2021izm, LIGOScientific:2023bwz,Janquart:2023mvf}.

Multiple images of CBC events can be searched for with a variety of methods
\citep{Haris:2018vmn,McIsaac:2019use,Li:2019osa,Liu:2020par,Janquart:2021qov,Lo:2021nae,Goyal:2021hxv,Janquart:2023osz,Ezquiaga:2023xfe,Li:2023zdl,Magare:2024wje}.
Effects that modify the waveforms,
including microlensing, millilensing, and type-II strongly lensed images,
can also be detected from a single signal directly,
provided the effect is large 
enough~\citep{Wright:2021cbn,Janquart:2021nus,Wang:2021kzt,Vijaykumar:2022dlp,Liu:2023ikc}.
In the particular case of a lensed BNS signal, it could also be 
identified as lensed by inspecting the measured tidal deformability, 
as the source masses and tidal parameters inferred without accounting for lensing would become 
incompatible for realistic EoSs~\citep{Pang:2020kle}.

\subsection{Strong Lensing}\label{sec:strong_lensing}

Most of the discussion in this paper concerns the strong lensing scenario,
hence we give here some more details on this regime.
In the frequency domain, the GW waveform of the $i^{\text{th}}$ 
lensed image ($h_{\mathrm{L}}^{i}$) is related to that of the 
original unlensed signal ($h_\mathrm{NL}$) as 
\begin{equation}
 \label{eq:strong-lensing-h}
    h_{{\mathrm{L}}}^{i}(f; \boldtheta, \boldphi_i) = \sqrt{\mu_{i}}\, h_{\mathrm{NL}}(f; \boldtheta)\, e^{(2i\pi f \Delta t_{i} - i\pi n_{i} \mathrm{sign}(f))},
\end{equation}
where $\boldtheta$ are the usual binary parameters, and 
$\boldphi_i = \{\mu_i, \Delta t_i, n_i\}$ are the lensing parameters 
for image $i$, with the magnification $\mu_i$, the time delay
$\Delta t_i$, and the Morse factor $n_i$~\citep{Dai:2017huk,Ezquiaga:2020gdt}.
The latter
can only take three values:
$n_i = 0, 0.5, 1$, corresponding to so-called image types I, II, and III,
respectively~\citep{Ezquiaga:2020gdt}.
Only type-II images can
potentially leave a detectable imprint on lensed
GW signals~\citep{Ezquiaga:2020gdt, Wang:2021kzt,Janquart:2021nus,Vijaykumar:2022dlp}.

From Eq. \eqref{eq:strong-lensing-h}, we see that some lensing
parameters are degenerate with the source parameters.
In particular, 
the amplitude of the unlensed signal is proportional to $1/D_L$,
which is a simple prefactor like the magnification. Additionally, 
the time of arrival of a lensed signal is $t_c + \Delta t_i$,
combining the usual time of coalescence, $t_c$, and the lensing
time delay. So, both the magnification and the time 
delay are degenerate with the luminosity distance and the time of
coalescence, respectively.
When estimating these parameters
under the unlensed hypothesis, we actually measure
\emph{effective} parameters
\begin{align}\label{eq:effective_parameters}
    D_L^{\mathrm{eff}} &= \frac{D_L}{\sqrt{\mu_i}}\,, \\
    t_c^{\mathrm{eff}} &= t_c + \Delta t_i \, .
\end{align} 

With a single GW image and no extra information
(either external or from a distortion of the waveform), 
it is not possible 
to disentangle the lensing effect from the true distance to the source
and its intrinsic properties
in all generality.
However, when several images are observed, it is possible to do 
\emph{lens reconstruction} to find the parameters of the lens--source system
and infer the time delay and magnification of each 
image~\citep{Hannuksela:2020xor,Wempe:2022zlk,Wright:2023npv,Seo:2023rjd,Poon:2024zxn}.

If not accounted for, lensing can also lead to biased
source parameter estimation. There is a general
redshift--mass degeneracy in CBC signals because we observe the redshifted signal
and have a direct GW measurement of the luminosity distance but not of the
redshift. 
Therefore, the observed detector-frame mass, $m^{\mathrm{det}}$,
for a binary component is
\begin{equation}\label{eq:det_frame_mass}
    m^{\mathrm{det}} = (1+z_s)\,m^\mathrm{src}\,,
\end{equation}
with $z_s$ the redshift of the source and $m^\mathrm{src}$ the source-frame mass.
The inferred source-frame mass is
\begin{equation}\label{eq:inferred_source_mass}
    m^\mathrm{src}_{\mathrm{inf}} = \frac{m^{\mathrm{det}}_{\mathrm{measured}}}{1+z_s(D_{L,\mathrm{measured}})} \,,
\end{equation}
where $m^{\mathrm{det}}_{\mathrm{measured}}$ is the measured detector-frame mass,
and $z_s(D_{L,\mathrm{measured}})$ is the redshift inferred from the measured luminosity distance
and a cosmological model.

Therefore, if the luminosity distance is biased by lensing, the
inferred source-frame masses will also be biased, and we will infer 
a higher (lower) source frame mass than the real one for
a magnified (demagnified) signal. This also 
means that exceptionally high-mass BBH events could potentially be
explained through lensing~\citep{LIGOScientific:2021izm,Diego:2021fyd}, 
though no compelling evidence 
has been found for this so far~\citep{LIGOScientific:2021izm}.
Similarly, some BNS events could be identified as NSBH events with a 
mass-gap component~\citep{Bianconi:2022etr,Magare:2023hgs,Canevarolo:2024muf}.

\subsection{Lensed Binary Neutron Star Mergers}\label{subsec:bns_lensing}

When an observed CBC contains a neutron star, the GW signal also
carries imprints of matter effects.
For example, BNSs have complicated post-merger
behaviors~\citep{Bauswein:2012ya,Bauswein:2014qla,Bernuzzi:2015rla,Tsang:2019esi} 
while the inspiral already is affected by the deformation of the neutron stars
due to tidal forces~\citep{Hinderer:2009ca,Damour:2009vw} 
and due to their own rotation~\citep{Laarakkers:1997hb,Poisson:1999ke,Harry:2018hke}.
This tidal deformation can be used to
probe the internal structure
and EoS~\citep{Agathos:2015uaa,Samajdar:2019ulq,LIGOScientific:2018cki}
of neutron stars.
The deformation of a neutron star due to the
gravitational field of its companion is described by the tidal 
deformability~\citep{Flanagan:2007ix,Hinderer:2009ca}
\begin{equation}\label{eq:tidal_deformability}
    \Lambda = \frac{2}{3} k_2 \left(\frac{c^2 R}{G m}\right)^5,
\end{equation}
where $k_2$ is the Love number, $R$ is the neutron star radius,
and $m$ is its mass. To calculate the value of this parameter for a 
given mass, one needs to solve the
Tolman-Oppenheimer-Volkoff (TOV) equation assuming an EoS~\citep{Hinderer:2009ca}. 

In a BNS GW signal, the leading-order tidal effects are encoded in
\begin{equation}\label{eq:lambda_tilde_def}
    \tilde{\Lambda} = \frac{16}{13} \sum_{i = 1,2} \Lambda_i \frac{m_i^4}{M^4} \bigg( 12 - 11 \frac{m_i}{M}\bigg) \,,
\end{equation}
where $M = m_1 + m_2$ is the total mass. So, the best measured 
effect is a combination of the masses and tidal deformabilities.
The individual tidal deformabilities are dimensionless
and not affected by cosmological redshift.
In $\LambdaTilde$,
the redshift dependency of the masses cancels out in the ratios,
so it is not changed by redshift either
when converting between the detector and source frames.

To detect lensed BNSs, one can leverage this fact that the inferred masses
are biased by the lensing effect, while the tidal deformabilities are
not, and both quantities are related~\citep{Pang:2020kle}. 
When an event with measurable tidal deformability is unlensed,
then the deformability calculated from the inferred masses (for a given EoS)
should coincide with the measurement directly from the GW data.
On the other hand, if the event is lensed,
the inferred masses are biased towards higher values due to the image magnification, leading to 
lower calculated values of the tidal deformability.
Under sufficiently strong magnification,
there would be a noticeable mismatch between the calculated
and measured values,
showing the event is lensed.

Moreover, should one have an accurate measurement of the tidal deformability,
one can correct the recovered parameters
for the lensing effect by 
finding a magnification such that the measured and calculated tidal 
deformabilities match. 
Alternatively, one can leverage this effect 
to account for lensing and compare the obtained event characteristics 
under the lens hypothesis
with those expected from lensed BNS population simulations. 
This can be used to assess whether a given event is a lensed BNS 
candidate.
For this to work, we however need to
be able to measure the tidal deformability and have a sufficient lensing effect
on the parameters. The former requires a loud signal,
and the latter requires relatively
high magnifications~\citep{Pang:2020kle}.

%% file: rates.tex
In this section,
we first discuss general predictions for the rates of lensed BNSs,
and then focus on the case of high magnifications,
as would likely be needed to explain GW230529 as a lensed BNS event.

\subsection{Rates from different population models}\label{subsec:rates_models}

Based on our expectations for source (CBC) and lens populations, we can
calculate the detection rate for strongly lensed CBCs
\citep{Ng:2017yiu,Li:2018prc,Oguri:2018muv,Xu:2021bfn,Mukherjee:2021qam,Wierda:2021upe,Magare:2023hgs}.
In particular, we need to assume some mass and redshift distributions
for the CBCs, a mass profile model and redshift distribution for the lenses,
and a strong lensing optical depth.
Calculating the rate of lensed events
requires simulating a large number of sources and lenses
and then evaluating the rate of detectable lensed events  \citep{Phurailatpam:2024enk} as 
\begin{equation}\label{eq:rate_calculation}
\begin{split}
    \mathcal{R}  = \mathcal{N}^{L}\langle P \left( {\rm obs}| z_{s}, \theta, \theta_{L}, z_{l}, \beta, {\rm{SL}}\right) & \rangle_{z_{s}\in P(z_{s}|{\rm SL}),\,  z_{l}\in P(z_{l}|z_{s}, {\rm SL}), \, \theta \in P(\theta),} \\
    & _{\theta_{L}\in P(\theta_{L}|z_{l}, z_{s}, {\rm SL}), \, \beta \in P(\beta|z_{s}, z_{l}, \theta_{L}, {\rm SL})}
\end{split}
\end{equation}
where $P(\mathrm{obs}|\dots,\mathrm{SL})$ is the probability of a strongly lensed (``SL'') event being observed and $\langle\dots\rangle$ indicates integrating it over the priors of all the listed dependencies.
Also, $\mathcal{N}^{L}$ is the normalising factor
\begin{equation}
    \mathcal{N}^{L} = \int_{z_{min}}^{z_{max}}R(z_{s})P({\rm SL}|z_{s})\frac{1}{(1+z_{s})}\frac{dV_{c}}{dz_{s}}dz_{s} ,
\end{equation}
$z_{s}$ is the source redshift, $z_{l}$ is the redshift of the galaxy lens, $\theta$ defines the source parameters, $\theta_{L}$ defines the lens parameters, and $\beta$ denotes the source position. $R(z_{s})$ is the merger rate density distribution regardless of whether the source is detectable or not, $P({\rm SL}|z_{s})$ is the probability of strong lensing for a source at redshift $z_{s}$, and $dV_{c}/dz_{s}$ is the differential co-moving volume at redshift $z_{s}$.

We also note that this $P(\mathrm{obs}|\dots,\mathrm{SL})$ depends
on the SNR of each event and on a threshold SNR ($\mathrm{SNR}_{\mathrm{cut}}$). 
The latter is used as a proxy for detectability,
and a value of 8 is a common approximation,
though see~\citet{Essick:2023lde} for a more subtle discussion.

To have a sense of the dependence of the expected rates on the input models,
we look at three different source distribution models:
\begin{enumerate}
\item a uniform mass distribution between 1 and 3 $M_{\odot}$
(a rather conservative upper limit for neutron star masses) with the merger-rate density from~\cite{2021ApJ...921..154W},
\item a bimodal Gaussian mass model from~\citet{Wysocki:2018mpo} with the merger-rate density from~\cite{2021ApJ...921..154W}, 
\item and the mass model and merger-rate density from the population synthesis run M43.A of~\citet{Belczynski:2017gds}\footnote{
We also investigated multiple other models from~\citet{Belczynski:2017gds},
of which M43.A is the one with the highest local merger-rate density
among those compatible with the latest LVK observed rates~\citep{KAGRA:2021duu}.
(See Table 3 in~\citet{Belczynski:2017gds} for details.)
}.
\end{enumerate}
For the simulations with all three of these populations,
we present results assuming
singular isothermal ellipsoid (SIE) lenses~\citep{Koopmans2009}.
We also performed comparisons with singular isothermal spheres \citep[SIS,][]{Witt:1990},
finding broadly consistent results.

Differences in these three setups can lead to different results on rates 
both directly and through causing different selection biases.
For example, models with lower source
masses may lead to higher magnifications because
the events needs a stronger boost from lensing to be detectable
for a given distance.

For all models, we perform simulations and calculate the rates
using the \textsc{LeR} package~\citep{Phurailatpam:2024enk}.
We assume a single LIGO-Livingston
detector with the noise power spectral density (PSD)
from the GW230529 LVK data release~\citep{ligo_scientific_collaboration_2024_10845779}.
We require only one of the 
lensed images to pass the SNR threshold of 8, since only one event
with properties similar to those for GW230529 was observed.
(See Section~\ref{subsec:multi_images} for a check on this assumption.)

\begin{table}
    \centering
    \begin{tabular}{cccc}
        \hline
        Population & Unlensed rate & Lensed rate & Relative rate \\
        & [$\rm yr^{-1}$] &  [$\rm yr^{-1}$] & \\
        \hline
        Uniform mass & $7.1 \times 10^{-1}$ & $1.5 \times 10^{-3} $ & $2.1 \times 10^{-3}$\\
        Bimodal Gaussian & $3.9 \times 10^{-1}$ & $7.0 \times 10^{-4}$ & $1.8 \times 10^{-3}$\\
         M43.A & $1.6$ & $2.0 \times 10^{-4}$ & $1.2 \times 10^{-4}$ \\
        \hline
    \end{tabular}
    \caption{
    Unlensed detected event rates, lensed detected event rates, and relative lensed rates
    from our various \textsc{LeR}-based simulations with different mass distributions and merger-rate functions.  
    \label{tab:rates_len_unlen}
    }
\end{table}

While some variations (of a factor of a few) are seen in the relative lensing 
rate from one model to another,
our calculations all agree that
the relative detection rate of lensed versus unlensed events
is smaller than $\approx 2\times 10^{-3}$,
see Table~\ref{tab:rates_len_unlen}.
Hence, from prior expectations alone, it is
unlikely for an event like GW230529 to be lensed.

\begin{figure}
    \includegraphics[keepaspectratio, width=0.49\textwidth]{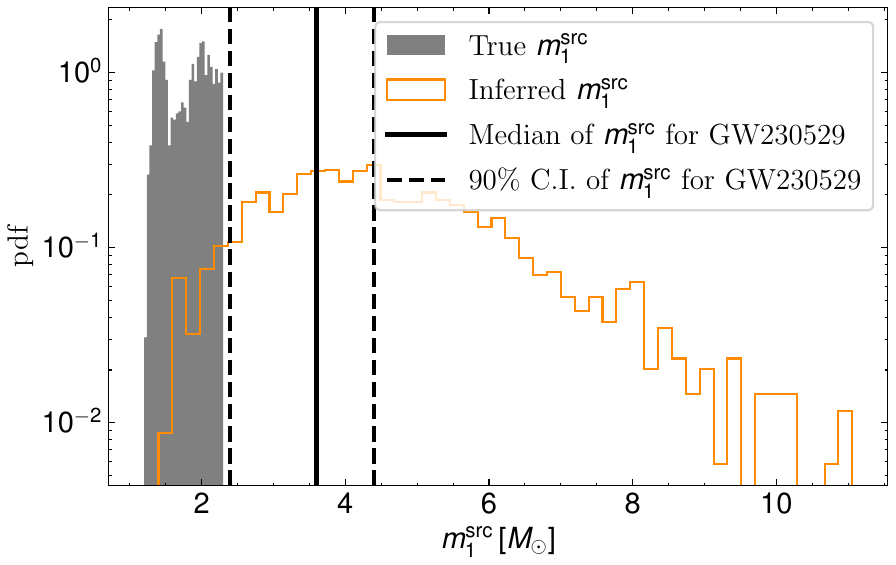}
    \caption{Example of how BNSs
    can appear as NSBHs
    if they are lensed but analyzed under the unlensed hypothesis.
    Here we show a bimodal mass distribution of true source-frame primary component masses.
    We compare it with the distribution one would infer from observations of lensed events from this population without accounting for lensing.
    Neglecting the lensing effect can lead to overestimated masses, pushing neutron stars into the 3--5\,$M_\odot$ mass gap. 
    Overlaid are the
    median and 90\% credible interval (C.I.) bounds for GW230529~\citep{LIGOScientific:2024elc}. 
    }
    \label{fig:bias_lensing_mass1}
\end{figure}

From the population simulations that go into these rate calculations,
we can also extract information about expected properties
of the lensed population.
This is shown in Fig.~\ref{fig:bias_lensing_mass1}, 
generated from the simulation with model (ii).
As has already been shown in other works~\citep{Bianconi:2022etr,Magare:2023hgs,Canevarolo:2024muf},
this demonstrates that for many lensed BNSs,
the heavier component object appears to fall into the mass gap
if source masses are inferred without accounting for lensing.
Nevertheless, the relative rates found for this and the other models
show that it is very unlikely to observe such
a lensed BNS event at O4a sensitivity.

\subsection{Probability of high-magnification BNS events}\label{subsec:magnification_proba}

The \textsc{LeR} rate calculations
work by simulating a large number of lensed and unlensed events,
then testing their detectability.
We can thus also obtain the expected distribution of the
lensing parameters for detectable lensed events.
In particular, we can see how often
detectable lensed BNSs would be magnified enough that they can be confused
with mass-gap events.
For simplicity, here we limit source redshifts to \mbox{$z_{\max} = 2$}.
This is sufficient because the detection horizon for unlensed BNSs is low,
so that detectable lensed events from even higher redshifts
would require even higher magnifications than what is needed for GW230529 to be
shifted into the BNS range.
(This will be demonstrated explicitly in Section~\ref{subsec:delensing}.)

\begin{figure}
    \centering
    \includegraphics[keepaspectratio, width=0.5\textwidth]{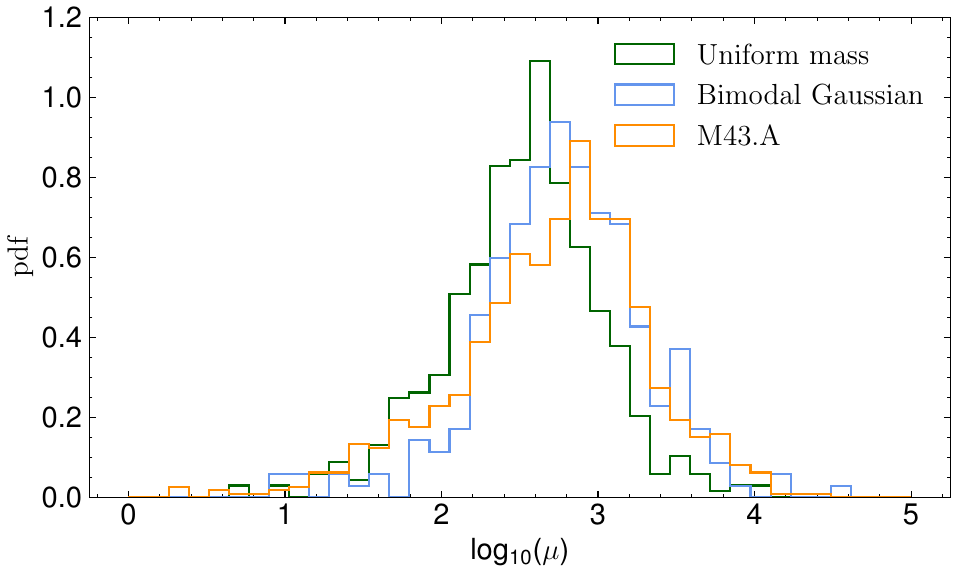}\\
    \caption{Histograms of the expected magnifications for detectable lensed BNS signals for the three simulations in our analysis.
    All three population models, representing different source mass models and merger-rate density functions,
    predict similar distributions.}
    \label{fig:distribution_magnifications}
\end{figure}

The magnification distributions
obtained for the three models are represented in Fig~\ref{fig:distribution_magnifications}.
We see that different mass and merger-rate density models predict similar distributions
with high probabilities for magnification in the hundreds to thousands.
Notably, these are the expected magnifications for detectable events only,
\emph{i.e.} they include selection effects.
Across all lensed events (detectable or not),
the standard result for the high-magnification limit
is $p(\mu) \propto \mu^{-3}$  \citep{Schneider:1992bmb}.
For example, this would make a magnification of $\mu=10^3$
about $10^3$ times less likely compared to $\mu=10^2$,
but due to events with higher magnification being detectable from further away
and hence from a larger source population,
this trend is significantly shallower in our results.

\begin{table}
    \centering
    \begin{tabular}{cccc}
        \hline
        Magnification & Uniform mass & Bimodal Gaussian & M43.A \\
        \hline
        $\geq 10^2$ & 86.6\% & 93.8\% & 87.2\% \\
        $\geq 10^3$ & 14.1\% & 31.4\% & 31.9\% \\
        $\geq 10^4$ & 0.4\% & 1.1\% & 0.9\% \\
        \hline
    \end{tabular}
    \caption{
    Probability to observe a magnification larger than a given threshold value
    in a detectable lensed BNS event,
    according to our various \textsc{LeR}-based simulations
    with different mass distributions and merger-rate functions.
    In all cases,
    the percentage of detectable events with magnification $\geq10^3$
    is quite substantial,
    while magnifications $\geq10^4$ are rare.
    \label{tab:magnification_probabilities}
    }
\end{table}

Based on these simulations,
we can check how likely it is to observe highly magnified lensed BNSs
appearing similar to the GW230529 observation.
We have used the obtained samples to calculate the probability of observing magnifications
at or above a certain threshold.
The values are shown in Table~\ref{tab:magnification_probabilities}.
There is a slight difference depending on the source population considered,
but the numbers are broadly in agreement:
The bulk of the distributions is below $\geq10^3$,
but values $\geq10^3$ are still quite plausible for all models,
while values $\geq10^4$ become rare in all simulations.

%% file: population.tex
While it is believed that the evolution of single stars cannot readily produce compact objects between $3-5 M_{\odot}$
\citep{Ozel:2010kle, Farr:2011bde},
there are other channels that can partially populate the purported mass gap,
such as a hierarchical merger of a BNS merger remnant with another neutron star.
In fact, previous GW observations had already suggested that the mass gap might not be completely empty \citep{KAGRA:2021duu}.
\cite{LIGOScientific:2024elc} provides a more detailed discussion on the possible formation mechanisms for GW230529.

With lensing, if the true source-frame mass of the primary component of GW230529 is less than what was inferred
and away from the mass gap because of the magnification bias,
then the binary system would indeed be more consistent with both prior GW observations and theoretical expectations.
However, this alternative explanation would require lensing of the GW signal with a certain magnification, which is by itself rare.
Here, we discuss this trade-off between population consistency and the probability of the lensing hypothesis,
first qualitatively and then quantitatively by calculating a corresponding Bayes factor.

\subsection{Qualitative considerations}
\label{sec:pop-qualitative}

To understand whether any gain in consistency with prior expectations by making the source lighter overall would be enough
to outweigh the penalty of invoking a rare phenomenon such as lensing,
we first need to consider a specific population model.
Both for the lensed and unlensed scenarios,
we assume that the overall CBC source population follows the \textsc{Power Law + Dip + Break} model \citep{Fishbach:2020ryj,Farah:2021qom},
with its maximum-likelihood parameters as reported in \citet{KAGRA:2021duu}.
The joint probability distribution of the component masses $p_{\rm pop}(m_1^{\textrm{src}}, m_2^{\textrm{src}})$ from this population model is factorized as
\begin{equation}
p_{\rm pop}(m_1^{\textrm{src}}, m_2^{\textrm{src}}) \propto p_{\rm pop}(m_1^{\textrm{src}}) p_{\rm pop}(m_2^{\textrm{src}}) \left( \dfrac{m_2^{\textrm{src}}}{ m_1^{\textrm{src}}} \right)^{\beta} \Theta(m_1^{\textrm{src}} \geq m_2^{\textrm{src}}),  
\end{equation}
where $p_{\rm pop}(m_{1,2}^{\textrm{src}})$ consists of two broken power laws
(one for neutron stars and one for black holes, respectively)
and a dip to model the relative scarcity between the two regions
(the ``mass gap'', though it is not completely empty in this model).

\begin{figure}
\includegraphics[width=\columnwidth]{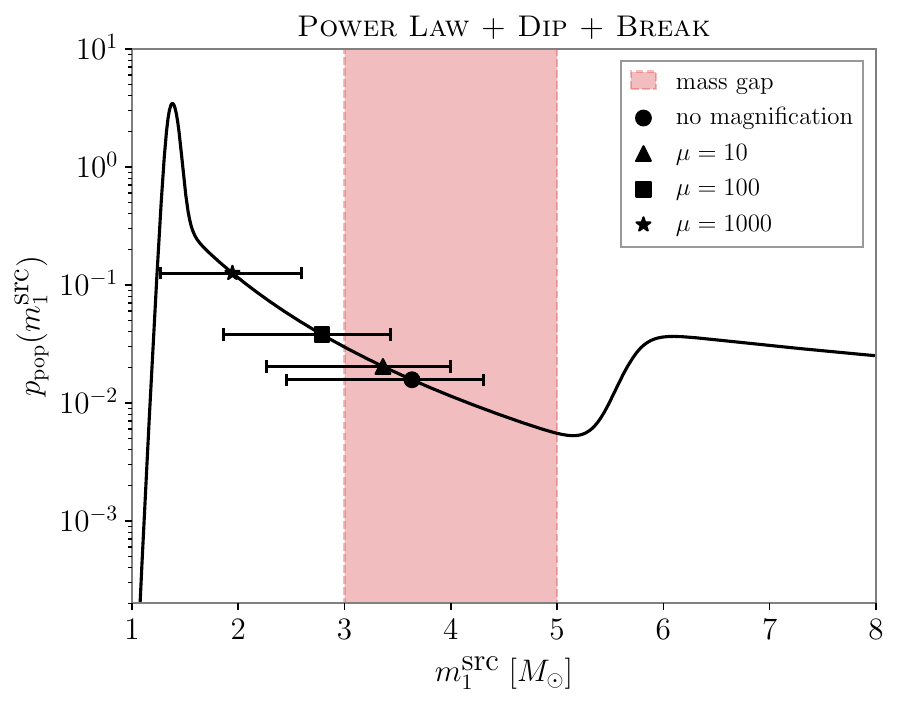}
\caption{\label{fig:PDB-m1}
Probability distribution of the source-frame primary mass $p_{\rm pop}(m_{1}^{\rm src})$
of CBC sources
as modeled by the \textsc{Power Law + Dip + Break} model.
The shaded band indicates the purported mass gap from 3--5\,$M_{\odot}$.
The median of the inferred $m_{1}^{\rm src}$ from \citet{LIGOScientific:2024elc}
without any magnification is shown as a dot,
with the bars showing the 90\% credible interval.
In addition, the median and the uncertainty on the source-frame primary mass assuming a magnification
of $\mu = 10$, $\mu = 100$ and $\mu = 1000$ are indicated by a triangle, a square and a star, respectively.
}
\end{figure}

In Fig.~\ref{fig:PDB-m1}, the probability distribution of the source-frame primary mass $m_{1}^{\rm src}$ from this model is shown.
Even before any further calculation, we see that there is some support for the primary of GW230529 to be outside the mass gap
\emph{without} assuming any magnification,
i.e. with the original inference from \citet{LIGOScientific:2024elc}.
In the same figure, we also indicate what the inferred source-frame primary mass would be
if the GW signal was magnified by some example magnification values of
$\mu = 10, 100, 1000$.
We see that with $\mu = 100$, more than half of the support moves outside of the mass gap,
and accordingly, the probability density is increased by a factor of $\approx 3$.
With $\mu = 1000$,
most of the support would be in the standard neutron star region,
outside the mass gap.
However, as discussed in Section~\ref{sec:rates},
higher magnifications are, in turn, less likely.

While a high magnification will push the primary to a lower mass
and hence a higher consistency with the expected population,
the mass of the secondary component,
receiving the same correction factor,
might be pushed towards the sub-solar region.
This can be problematic because the secondary
needs to be compact enough such that the two objects will not be in contact already
before the GW emission due to their inspiral reaches the sensitive frequency band of the LVK detectors ($\sim 10\,{\rm Hz}$).
The neutron star minimum mass is expected to be around $1M_{\odot}$ \citep{Suwa:2018uni}.
While white dwarfs can be lighter,
they do not have the necessary compactness and a neutron star--white dwarf binary will already be in contact
at around $\sim 0.2\,{\rm Hz}$~\citep{Golomb:2024mmt}.
Therefore, for a sub-solar secondary,
another exotic scenario such as a primordial black hole or other kinds of exotic compact objects will be needed.

\begin{figure}
\includegraphics[width=\columnwidth]{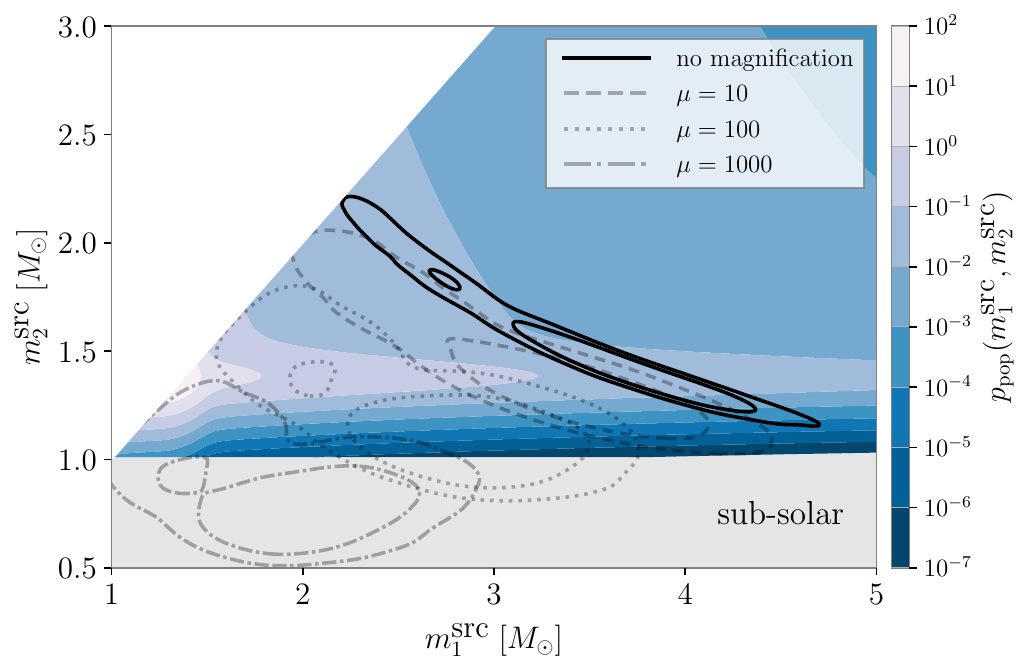}
\caption{\label{fig:PDB-m1-m2}
Joint probability distribution of the source-frame component masses $p_{\rm pop}(m_{1}^{\rm src}, m_{2}^{\rm src})$
as modeled by the \textsc{Power Law + Dip + Break} model.
As the magnification $\mu$ increases,
the inferred source-frame component masses decrease
(from the solid contours for $\mu = 1$ to the dash-dotted contours for $\mu = 1000$).
In particular, when $\mu = 1000$,
a large portion of the support for the secondary component mass
lies in the sub-solar ($<1 M_{\odot}$) region.
}
\end{figure}

We present joint probability distributions for both component masses in Fig.~\ref{fig:PDB-m1-m2},
both for the original results and under the same three example magnification values as before.
We see that starting from $\mu \sim 10$,
some of the support for the source-frame secondary mass $m_{2}^{\rm src}$ will be below $1\,M_{\odot}$.
For $\mu = 1000$,
the primary component will be outside the mass gap,
but most of the support for the secondary will be in the sub-solar region.
This aspect will be discussed more in depth in Section~\ref{subsec:delensing}.

\subsection{Bayes factor calculation}
\label{sec:pop-BF}

We now consolidate all these considerations,
both from the perspective of source population consistency
and the relative probability of lensing with a certain magnification,
into one single number---a Bayes factor $\mathcal{B}^{\rm L}_{\rm NL}$
between the lensed hypothesis $\mathcal{H}_{\rm L}$
and the not-lensed hypothesis $\mathcal{H}_{\rm NL}$.
Following \cite{Lo:2021nae}, it is defined as
\begin{equation}
	\mathcal{B}^{\rm L}_{\rm NL} \equiv \dfrac{p(d_{\rm GW}|\mathcal{H}_{\rm L})}{p(d_{\rm GW}|\mathcal{H}_{\rm NL})},
\end{equation}
where $p(d_{\rm GW}|\mathcal{H}_{\rm L})$ and $p(d_{\rm GW}|\mathcal{H}_{\rm NL})$ are the marginal likelihoods
for the observed data $d_{\rm GW}$
under each  hypothesis.
Considering only one possibly lensed signal,
the marginal likelihood under $\mathcal{H}_{\rm L}$ is given by
\begin{multline}
p(d_{\rm GW}|\mathcal{H}_{\rm L}) \propto \int {\rm d}z_s p_{\rm pop}(z_s|\mathcal{H}_{\rm L}) \left[ \sum_{n=0,1/2,1} \int {\rm d} m_1^{\rm src} {\rm d}m_2^{\rm src} {\rm d}\mu \right. \\
\left. p(d_{\rm GW}|m_{1,2}^{\rm det}=(1+z_s)m_{1,2}^{\rm src}, D_{\rm L}=\dfrac{D_{\rm L}^{\rm src}}{\sqrt{\mu}}, n) p_{\rm pop}(m_1^{\rm src}, m_2^{\rm src}) p(\mu) \right],
\end{multline}
where the term enclosed by the square brackets is the marginal likelihood $\mathcal{L}_{\mathcal{H}_{\rm L}}(z_s)$
as a function of the source redshift under $\mathcal{H}_{\rm L}$.
We will see in Section~\ref{subsec:type_II} that for this event,
we could not find any evidence for a type-II (i.e., $n=1/2$) lensed signal,
and therefore we will drop the summation over $n$ in the marginal likelihood.
Similarly, the marginal likelihood under $\mathcal{H}_{\rm NL}$ is given by
\begin{multline}
p(d_{\rm GW}|\mathcal{H}_{\rm NL}) \propto \int {\rm d}z_s p_{\rm pop}(z_s|\mathcal{H}_{\rm NL}) \left[ \int {\rm d} m_1^{\rm src} {\rm d}m_2^{\rm src} \right. \\
p(d_{\rm GW}|m_{1,2}^{\rm det}=(1+z_s)m_{1,2}^{\rm src}, D_{\rm L}=D_{\rm L}^{\rm src}) p_{\rm pop}(m_1^{\rm src}, m_2^{\rm src}) \Biggr ],
\end{multline}
and again the term enclosed by the square brackets is
the marginal likelihood $\mathcal{L}_{\mathcal{H}_{\rm NL}}(z_s)$
as a function of the source redshift under this hypothesis.

Using the $\texttt{hanabi.hierachical}$ code~\citep{Lo:2021nae}
and assuming $p(\mu) \propto \mu^{-3}$ \citep{Schneider:1992bmb}\footnote{Selection effects are accounted for in the normalization of the marginal likelihoods. Therefore, the prior for the magnification $p(\mu)$ to be used in the Bayes factor calculation should be the one \emph{without} the selection effects, which is different from the expected distribution, e.g., in Fig.~\ref{fig:distribution_magnifications}.},
we compute the Bayes factor for GW230529 to be $\approx 1:58$ \emph{disfavoring} the lensed hypothesis.
This is consistent with intuition from the overall lensing rates and
from the illustrative Figs.~\ref{fig:PDB-m1} and~\ref{fig:PDB-m1-m2}:
the gain in consistency with population expectations from a lighter primary is not sufficient
to overcome the penalty from needing the GW signal to be lensed with a sufficient magnification.
Combined with the prior odds for the lensed versus the not-lensed hypothesis,
related to the relative rate of lensed events estimated as $\lesssim2\times10^{-3}$ in Section~\ref{subsec:rates_models},
statistically speaking GW230529 is unlikely to be magnified significantly.

%% file: delensing.tex
\subsection{A lensed binary with a neutron star as the primary?}
\label{subsec:delensing}

To further investigate whether GW230529 could be a lensed BNS, we
can correct for the potential lensing effect, 
\emph{i.e.} find the source parameters assuming lensing,
and see whether this gives a plausible system. To do so,
we assume the primary to be a neutron star with a certain mass 
and see what would be the mass of the secondary, the source 
redshift, and the magnification. We can then verify whether these 
properties correspond to those expected in a lensed BNS system or
to something more exotic or even non-physical.

We use three sets of results from~\citet{LIGOScientific:2024elc}, all with the IMRPhenomXPHM waveform~\citep{Pratten:2020ceb} but with different priors:
\begin{description}
    \item[(i)] one with sub-extremal spin amplitudes of the two components, $\chi_{1,2} \leq 0.99$,
    called ``high spin'',
    \item[(ii)] one where the spin amplitude of the secondary is limited to $0.05$, $\chi_2 \leq 0.05$,
    called ``low spin secondary'',
    \item[(iii)] one where both spin amplitudes are limited below $0.05$, $\chi_{1,2} \leq 0.05$,
    called ``low spin both''.
\end{description}

\begin{figure}
    \centering
    \includegraphics[width=0.49\textwidth]{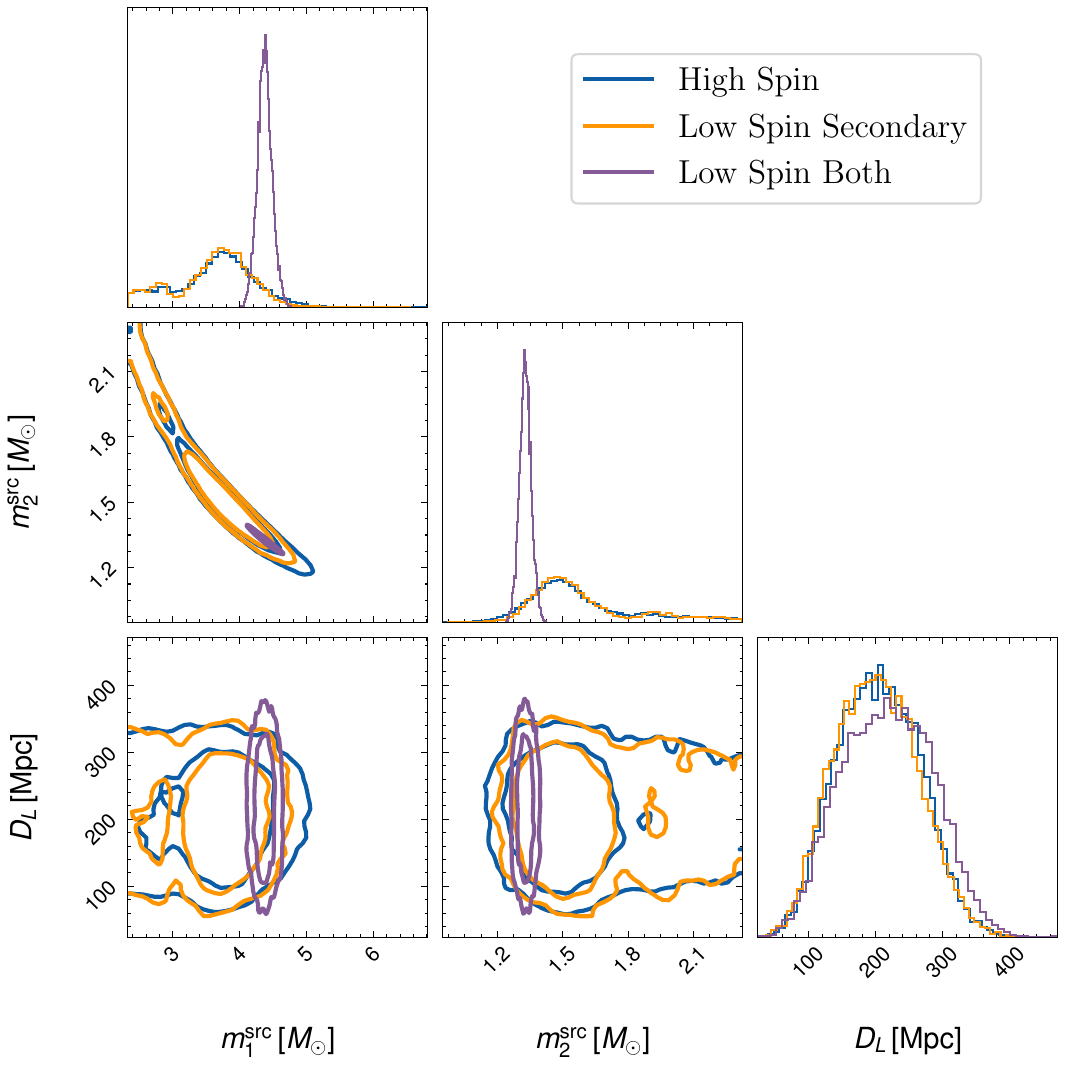}
    \caption{Posterior distributions for the source-frame component masses and the 
             luminosity distance for analyses from \citet{LIGOScientific:2024elc}
             done under the unlensed hypothesis, with the IMRPhenomXPHM 
             waveform and different spin priors.
             The unconstrained (``high spin'') prior and the one with
             a constraint only on the secondary object's spin (``low spin secondary'') give
             roughly the same results. The ``low spin both'' prior leads to
             a preference for more asymmetric masses.
             }
    \label{fig:corner_initial}
\end{figure}

In Fig.~\ref{fig:corner_initial}, we show the posterior distributions 
for the component masses
and luminosity distance inferred under the 
unlensed hypothesis,
with masses converted from the detector frame to the source frame using equation~\eqref{eq:inferred_source_mass}.
The ``low spin both'' prior leads to preferring more asymmetric masses,
while the two other results are broader,
with a preference for more symmetric masses, and similar to each other.

Under the hypothesis that GW230529 is lensed and the primary is a neutron star,
we can instead use the neutron star mass distribution to pick a maximum source mass $m_1^\mathrm{src}$,
corresponding to the least degree of lensing required under that hypothesis.
Now we can use equation~\eqref{eq:det_frame_mass} to compute the source redshift
from the measured detector-frame primary mass. 
We can also find the source-frame mass for the secondary from the measured detector-frame value, using the same redshift with the same equation.
Additionally, with
equation~\eqref{eq:effective_parameters}
we can compute the magnification the event would have undergone
for its masses to resemble those inferred for GW230529 under the unlensed hypothesis. 
Finally, assuming some EoS, it is then also possible 
to solve the Tolman-Oppenheimer-Volkoff equation and find the tidal deformabilities
for the two components.
Here, we use the maximum probability EoS from~\citet{Huth:2021bsp}.
Note that we use this regardless of the spin amplitudes considered for the event.

Using this procedure, we investigate what kind of system could
lead to observables similar to those of GW230529, under several
choices for $m_{1}^\mathrm{src}$:
first, we pick the median, the lower and the upper quantiles
at 90\% confidence for the maximum mass of a neutron star from~\citet{KAGRA:2021duu},
$1.99^{+0.29}_{-0.23}\,M_{\odot}$.
To be conservative, 
we also include the possibility of the highest neutron star mass 
being $2.5\,M_{\odot}$~\citep{Rocha:2023xwp,Fan:2024zqk}. 

\begin{figure}
    \centering
    \includegraphics[width=0.49\textwidth]{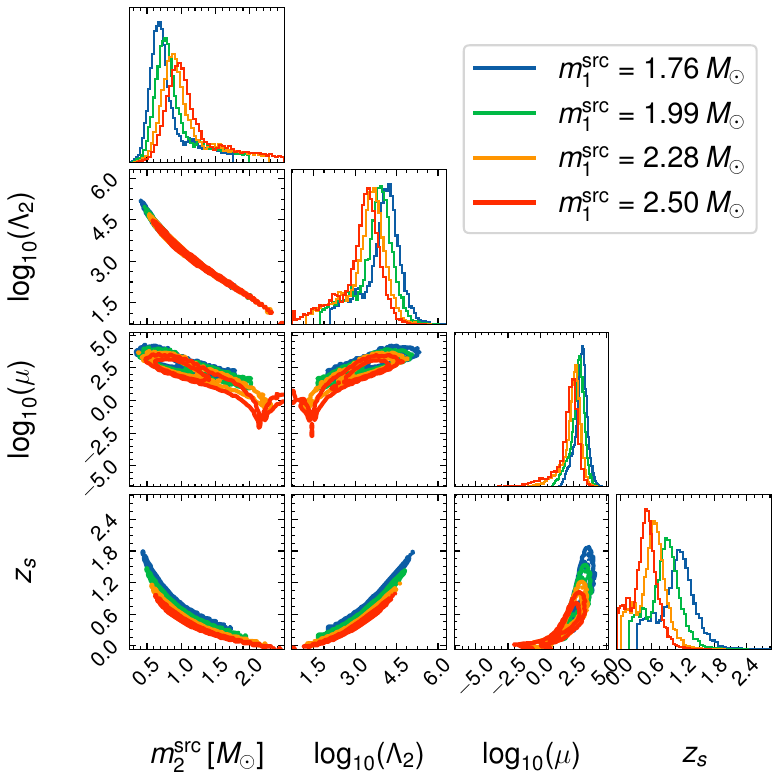}
    \caption{Posterior distributions under the lensed hypothesis for the inferred source-frame mass of the
    secondary object, its corresponding tidal deformability, as well as the magnification
    and source redshift. All results are obtained using the 
    ``high spin'' prior, but we assume different primary masses. 
    There is a non-negligible support for values in extreme
    regions (e.g. sub-solar mass, very high deformability)
    not matching the expectations for a lensed BNS
    or inconsistent with the GW230529 observation.
    \label{fig:corner_high_spin_delens}
    }
\end{figure}

Results for the agnostic ``high spin'' prior 
and the four masses are shown in Fig.~\ref{fig:corner_high_spin_delens}.
We find significant support
for the secondary being a sub-solar mass object,
which would be considered exotic~\citep{LIGOScientific:2022hai}.
Assuming the object is 
still made of nuclear matter, this leads to large tidal deformabilities.
One could expect these to be measurable,
which we will check further in the next two subsections.
Additionally, when looking at the expected lensing parameters, we 
find the biggest support for relatively large magnifications, 
with a significant weight 
above $\mu = 10^3$, except for $m_1^\mathrm{src} = 2.5 M_\odot$, where
only $\sim 10\%$ of the samples are above this limit.
This is still compatible with our expectations
from population models
but pushing into the less likely regime---see Fig.~\ref{fig:distribution_magnifications}. 
The source redshift would still be relatively low, corresponding
to low optical depth~\citep{Oguri:2019fix}, \emph{i.e.} a low lensing
probability.
In the end, only a limited fraction of the samples
could lead to a viable lens-source system not requiring measurable tidal deformabilities
or an exotic low-mass secondary object.

\begin{figure}
    \includegraphics[keepaspectratio, width=0.49\textwidth]{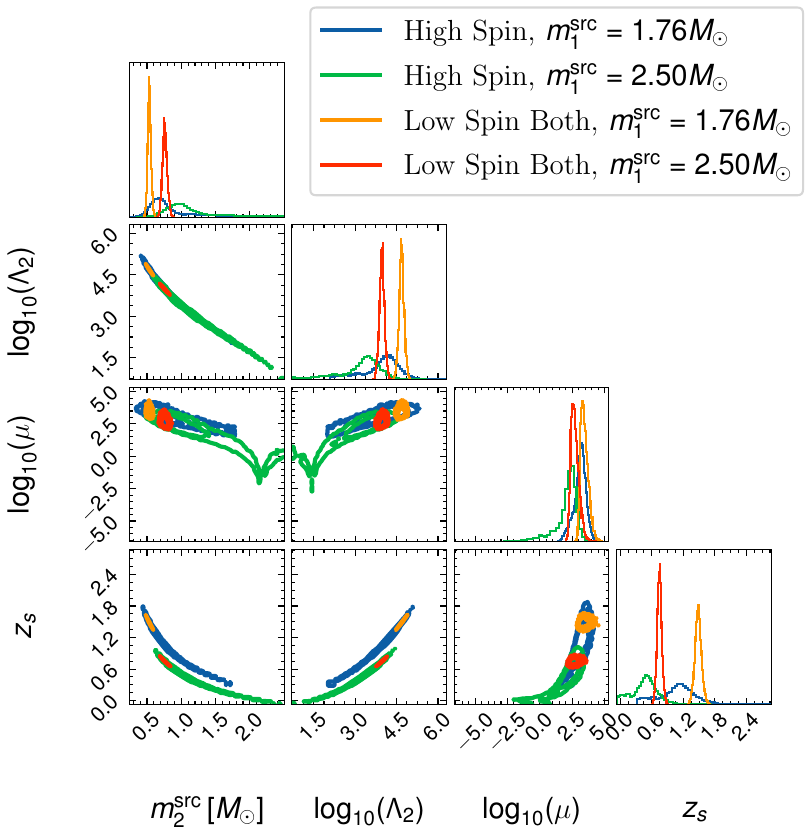}
    \caption{Posterior distributions under the lensed hypothesis
    for the ``high spin'' and ``low spin both'' priors,
    for the highest and lowest primary source mass considered before.
    The ``low spin both'' prior,
    designed to match expectations for BNSs,
    leads to more exotic results
    since the secondary would certainly be sub-solar, and the magnification much larger than expected.
    \label{fig:corner_compa_high_low_both}
    }
\end{figure}

The ``low spin secondary'' prior leads to the same conclusions,
though results are significantly different
for the ``low spin both'' prior, as shown in
Fig.~\ref{fig:corner_compa_high_low_both}.
For simplicity, we now show only the highest and lowest primary mass considered before.
With the ``low spin both'' prior, we always obtain a sub-solar mass secondary
and extremely high tidal deformabilities.
The magnifications would again be at the higher end of the distribution from our simulations,
with higher support above $\mu = 10^3$,
and also above $10^4$ for $m_1^\mathrm{src}=1.76\,M_\odot$.
Source redshifts still stay below 2,
matching the cut-off assumed in Section \ref{subsec:magnification_proba}.
Overall, imposing low spin constraints on both objects---as, 
in principle, expected for a BNS---leads to the most exotic results
least compatible with an actual lensed BNS.

We summarize the probability for the lightest object to be sub-solar in the various cases in 
Table~\ref{tab:sub_solar_proba}. For the ``low spin both'' prior, the secondary mass
is always below $1\,M_{\odot}$. If we assume the mass for a neutron star can 
be as low as $0.8\,M_{\odot}$, the probability for the lightest object to be lower
than this limit is 100\%, except for the highest $m_{1}^\mathrm{src}$ considered,
where it becomes $86.8\%$. So, if GW230529 was lensed and 
its spin amplitudes constrained below 0.05, then it would be nearly certain for 
the secondary to be a sub-solar mass object.
Such an object would be astrophysically unexpected
and either have very high tidal deformability or need to be a primordial black hole or
a beyond standard model physics object.
Therefore, what initially appears to be the most natural version of the BNS lensing 
scenario is unlikely to be consistent with an actual BNS
and instead requires the addition of another low-probability phenomenon.

For the other spin priors, the probability for the
secondary to be sub-solar is lower,
but still substantial.
While the system becomes 
less exotic in terms of masses, it would still require relatively high 
magnifications, and potentially large tidal deformabilities on the secondary
(if it is not a black hole)
to be compatible with the lensing hypothesis. Moreover, this corresponds
to the case where the spins do not match our expectations for BNSs.
So, should the event be lensed, it would still correspond to an 
astrophysically unexpected system.

\begin{table}
    \centering
    \begin{tabular}{cccc}
        \hline
        $m_1^\mathrm{src}$ ($M_{\odot}$) & High spin & Low spin secondary & Low spin both \\
        \hline
        1.76 & 67.9\% (50.8\%) & 67.4\% (51.1\%) & 100\% (100\%) \\
        1.99 & 60.2\% (34.2\%) & 60.8\% (33.8\%) & 100\% (100\%) \\
        2.28 & 46.5\% (16.3\%) & 46.8\% (14.5\%) & 100\% (99.9\%) \\
        2.5 & 33.3\% (8.5\%) & 32.7\% (6.7\%) & 100\% (86.8\%) \\
        \hline 
    \end{tabular}
    \caption{Probability for the secondary of a lensed GW230529 to be lighter
    than $1\,M_{\odot}$ (``sub-solar'' objects)
    or (in parentheses) $0.8\,M_{\odot}$
    as a more conservative limit for the lightest neutrons stars,
    for different prior choices from \citet{LIGOScientific:2024elc}.
    The ``low spin both'' prior leads to a
    quasi-certainty for the secondary to be sub-solar. 
    For the other priors, this probability decreases
    as we allow higher masses for the primary,
    but always stays substantial.
    \label{tab:sub_solar_proba}
    }
\end{table}

\subsection{Expected tidal deformabilities under the lensing hypothesis}\label{subsec:tidal_compatibility}

As explained in~\citet{Pang:2020kle}, if an event is lensed, there would be a 
discrepancy between the measured tidal deformability and the one expected from 
the source-frame masses found when undoing the lensing effects. 
Therefore, we now consider the tidal deformabilities
measured in~\citet{LIGOScientific:2024elc} from the runs with the
IMRPhenomPv2\_NRTidalv2 waveform~\citep{Dietrich:2019kaq}
and with two different priors:
one with a low-spin constraint on the secondary,
and one with a low-spin constraint on both objects.
We compare these measurements against the deformabilities we would expect
from the masses inferred under the lensing hypothesis.

After correcting for lensing as above,
we again calculate the tidal deformabilities for the two objects by
using the maximum probability EoS from~\citet{Huth:2021bsp},
and use equation~\eqref{eq:lambda_tilde_def}
to find the distribution for $\tilde{\Lambda}$. 
This parameter is the best measured combination of tidal deformabilities
since it appears at the lowest PN order in the
waveform~\citep{Hinderer:2009ca}.

\begin{figure*}
    \includegraphics[keepaspectratio, width=0.49\textwidth]{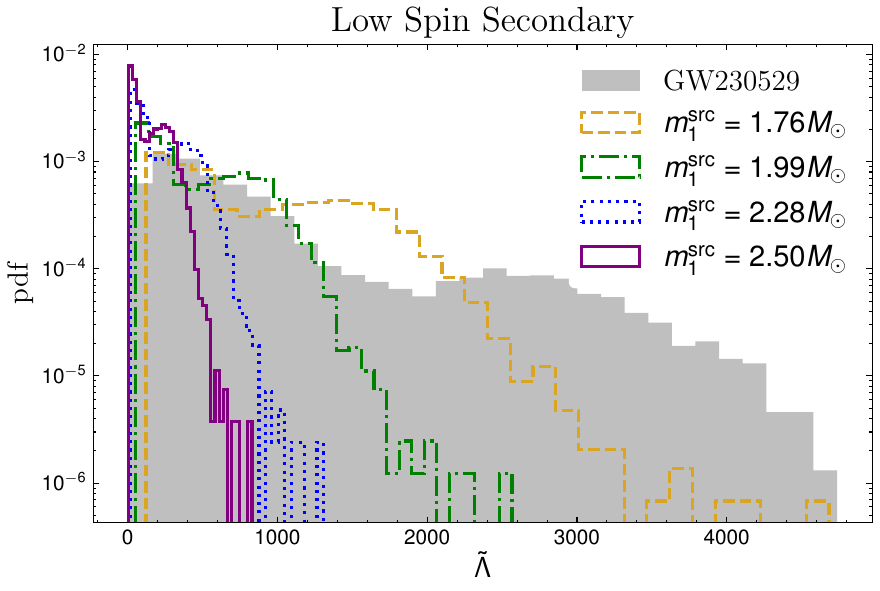}
    \includegraphics[keepaspectratio, width=0.49\textwidth]{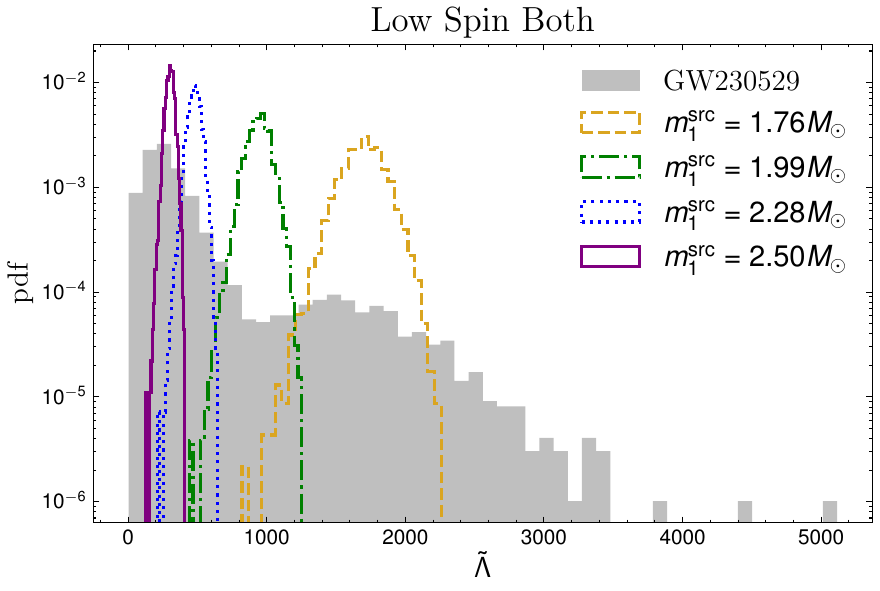}
    \caption{Comparison between the tidal deformability distributions expected for systems with GW230529's masses after correcting for lensing and the measured ones. 
    The left panel shows the results for the 
    low-spin constraint on the secondary, while the right panel shows the results for the 
    double low-spin constraint. The distributions are overlaid with the measured 
    $\tilde{\Lambda}$ values measured for each spin priors used when analyzing 
    the event with IMRPhenomPv2\_NRTidalv2. 
    \label{fig:tidal_comparisons}
    }
\end{figure*}

The left panel in Fig.~\ref{fig:tidal_comparisons} shows the 
$\tilde{\Lambda}$ distributions found for the ``low spin secondary'' prior with
the different $m_{1}^\mathrm{src}$ choices as before,
compared with the posterior found in the original analysis.
The right panel shows the same
for the ``low spin both'' prior.
In the first case, the distributions 
found by correcting for the lensing effects are broad
and span relatively low values,
due to the higher masses expected for the 
secondary in this case. For the highest $m_{1}^\mathrm{src}$ choices, $\tilde{\Lambda}$
would be the lowest, which would likely lead to measurements
that cannot exclude zero deformability.
For $m_1^\mathrm{src} < 2\,M_\odot$,
the range of possible values increases and part of the posterior
would lead to measurable tidal deformabilities, contrary to what is seen. 
Still, in all cases, there is support for tidal deformability values 
compatible with the measured posteriors and with zero.

For the ``low spin both'' prior, shown in the right panel of Fig.~\ref{fig:tidal_comparisons},
the distributions are more concentrated
towards higher values,
so that for $m_1^\mathrm{src} < 2\,M_\odot$ they begin to
fall mostly in the lower density regions of the measurements.
Therefore, should we have such tidal deformabilities, the highest mass cases 
would probably lead to a small enough $\tilde{\Lambda}$ to be unmeasurable.
On the other hand, for the lowest
$m_1^\mathrm{src}$ considered, most of the corresponding tidal deformabilities
are high enough to be measurable and would lead to posteriors not compatible
with zero, contrary to the observations from the GW230529 analyses.

\subsection{Injection Studies}\label{subsec:injections}

Going further, we can study with simulated signals if the high tidal deformability
that a lensed low-mass secondary would present
should have been observable in GW230529.

\begin{figure*}
    \centering
    \includegraphics[keepaspectratio, width=0.49\textwidth]{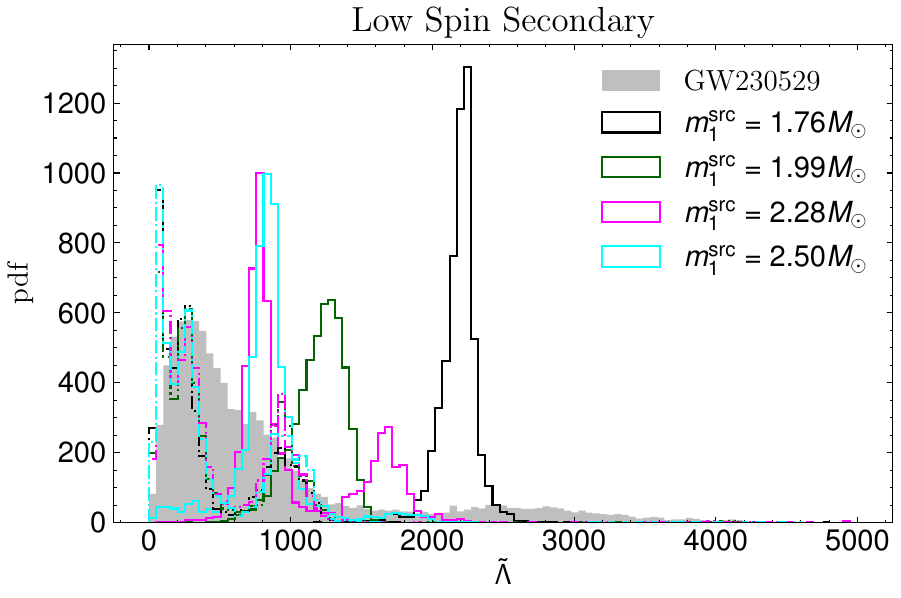}
    \includegraphics[keepaspectratio, width=0.49\textwidth]{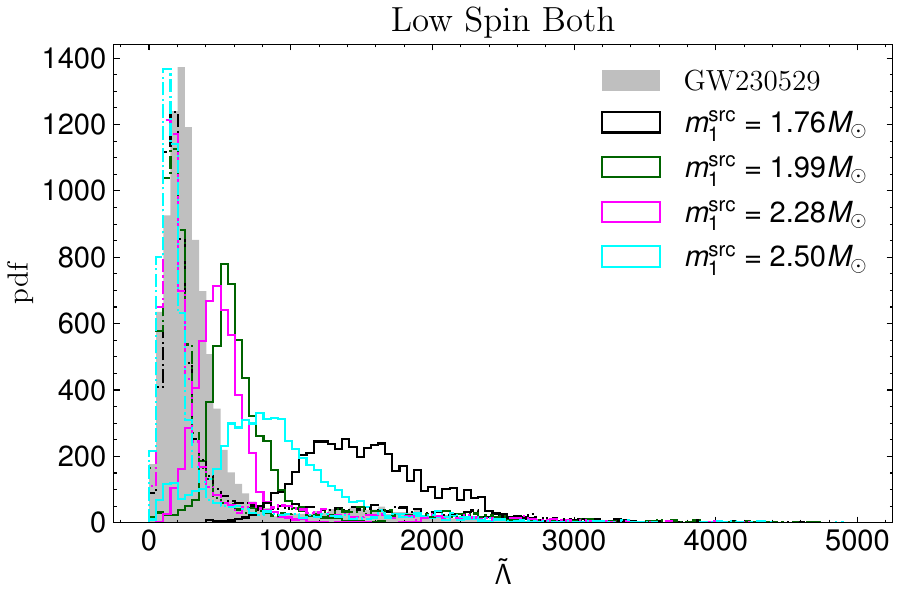}
    \caption{Posteriors for the $\tilde{\Lambda}$ parameter for maximum likelihood injections of the 
    lensing-corrected parameters (solid lines),
    injections with zero tidal deformability for the secondary object (dot-dashed lines), 
    and those from the parameter estimation runs for GW230529 with the same spin priors.
    The left panel shows the results for the ``low spin'' prior only on the secondary component,
    while the right panel shows the results for a ``low spin both'' prior.
    For both cases, the posterior when fixing $m_1^\mathrm{src}=2.5\,M_{\odot}$ is still compatible with zero,
    although peaked away from that value. This compatibility with zero becomes much weaker 
    if for lower $m_1^\mathrm{src}=2.28\,M_{\odot}$, and $\tilde{\Lambda}=0$ is excluded for the other
    assumed values.
    Meanwhile, for both priors the actual GW230529 posterior has support for zero,
    indicating that a lensed system with a primary of more conventional neutron star mass
    should have led to a different observation.
    \label{fig:tidal_deformabilities_rec}
    }
\end{figure*}

In Section~\ref{subsec:delensing} we have seen that, when correcting for the lensing effects, the
secondary object tends to have low masses.
Another implication
is that $\tilde{\Lambda}$ is likely to be high for mergers containing such objects,
since lower neutron star masses lead to higher tidal deformabilities.
In a more general simulation study,
\citet{Golomb:2024mmt} found that if both components are neutron stars
and at least one is sub-solar, this should be clearly measurable.
So, should the tidal deformability be measurable while the inferred source-frame component masses 
are relatively high, it would hint towards a possible lensed detection
with overestimated masses~\citep{Pang:2020kle}.

We now take the maximum likelihood parameters from the 
results derived in Section.~\ref{subsec:tidal_compatibility} when accounting for lensing for the 
IMRPhenomPv2\_NRTidalv2 waveform and inject them into Gaussian noise generated 
from the power spectral density for LIGO Livingston when GW230529 was observed \citep{ligo_scientific_collaboration_2024_10845779}.
Then, we analyze them
using the relative binning code from~\citet{Narola:2023men} and verify whether the recovered tidal
posteriors are more informative than those obtained on the real data.
Both injection and recovery are done with the TaylorF2 waveform \citep{Mishra:2016whh}
to avoid issues due to tapering faced with more complete 
models~\citep{Wouters:2024oxj}.
Moreover, BNS signals are dominated by the inspiral phase, where this
waveform model matches well with other waveform models, such as those from the Phenom family. 
For reference, we also do the same analysis for an injection with zero tidal deformability.

The $\tilde{\Lambda}$ posteriors for the various injections are shown in Fig.~\ref{fig:tidal_deformabilities_rec}.
The left panel shows results based on the ``low spin secondary'' prior,
and the right panel for the ``low spin both'' prior. Since the tidal deformabilities
of both components are larger for smaller $m_1^\mathrm{src}$,
the posteriors become more informative as we go down in $m_1^\mathrm{src}$. From our injections,
it seems like a posterior mostly consistent with zero is possible only if the primary is a black hole or 
a heavy neutron star with $m_1^\mathrm{src} > 2.3\,M_{\odot}$.
On the contrary, the actual GW230529 posteriors for $\tilde{\Lambda}$ peak close to zero and 
are broadly compatible with this value, meaning they are not compatible with lower masses and more magnified events.
Moreover, when doing the injections without tidal deformabilities, the recovered posteriors are consistent with zero 
and much closer to those for GW230529.

We note here that our injections were done only for the maximum likelihood values and a single EoS.
More in-depth studies covering the entire posterior support and various EoSs would help
increase our confidence in the conclusions from a non-measurable $\tilde{\Lambda}$ parameter. Still, 
our experiments indicate that the source is unlikely to have large tidal deformabilities and is closely
consistent with the standard interpretation of
an unlensed binary composed of a black hole and a neutron star, where one could also 
not measure $\tilde{\Lambda}$.

%% file: extra_checks.tex
Beyond considerations of population consistency and the observability of tidal deformations
under the lensing scenario,
it is also worth checking for direct signatures of lensing in the data of GW230529 and around it,
using some of the methods previously developed for LVK lensing searches~\citep{LIGOScientific:2021izm,LIGOScientific:2023bwz}.

\subsection{Multiple Images}\label{subsec:multi_images}

If a GW is strongly lensed, then we expect to be able to
detect additional lensed GW images at different arrival times,
which could range from minutes to years,
depending on the mass and characteristics of the lens.
In particular, we expect a highly magnified GW to have lensed counterparts
with even shorter time delays,
of the order of seconds to days
(for example, see \cite{Smith:2022vbp} in the context of lensed BNS systems), and even overlapping with each other \citep{Lo:2024wqm}.

Regardless of the prior expectations on strong lensing discussed in the previous sections,
it is still useful
to try and search for potential lensed GW counterparts of GW230529 in the data.
Given that only 4096 seconds of data, centered around the time of the event,
have been released for now,
it is very challenging to perform the usual sub-threshold targeted searches
\citep[e.g.,][]{McIsaac:2019use, Li:2019osa, Li:2023zdl}.
These would require a longer observation time to accumulate background triggers
(due to noise)
for proper assignment of statistical significance, 
such as a false-alarm rate,
to triggers found by those searches.

\begin{figure}
\centering
\includegraphics[width=\columnwidth]{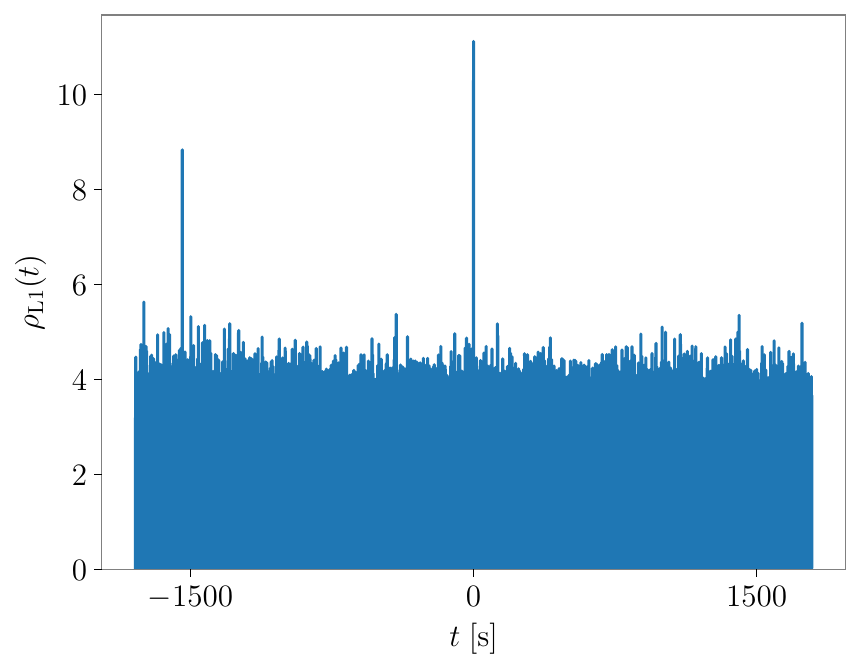}
\caption{\label{fig:pycbc_snr_timeseries}
The SNR time series when filtering the 4096\,s of LIGO Livingston data around GW230529
against the maximum-likelihood waveform
found by the parameter estimation analysis from \citet{LIGOScientific:2024elc} using the IMRPhenomXPHM waveform model.
The peak at time $t = 0$\,s corresponds to GW230529.
The secondary peak at $t \approx -1500$\,s,
fails the signal consistency test in \citet{Allen:2004gu} and hence is inconsistent with a genuine GW signal.
}
\end{figure}

Here, we opt to instead simply matched-filter the limited public data against a single template,
as a \emph{preliminary analysis}.
Fig.~\ref{fig:pycbc_snr_timeseries} shows the SNR time series,
computed using the \texttt{PyCBC} package\footnote{
Not to be confused with the full \texttt{PyCBC} pipeline that uses the same codebase.
Here we did not run the full search pipeline.
},
when filtering the LIGO Livingston data against the maximum-likelihood waveform
found by the parameter estimation analysis from \citet{LIGOScientific:2024elc} using the IMRPhenomXPHM waveform model \citep{Pratten:2020ceb}.
We have also used \texttt{GstLAL} to compute the SNR time series,
and the result is consistent with the one shown here.
The peak at time $t = 0$\,s corresponds to GW230529.

There is a secondary peak, with a matched-filter SNR of $\approx 9$ at $t \approx -1500$\,s.
This does not correspond to a genuine astrophysical signal 
since the corresponding trigger fails the signal consistency test in \citet{Allen:2004gu},
with a reduced chi-square $\chi_{\rm r}^2 \gg 1$.
Therefore, it is extremely inconsistent with a genuine GW
and is likely due to non-Gaussian noise fluctuations (\emph{i.e.} instrumental glitches).
After discarding this spurious peak in the SNR time series at time $t \approx -1500$ s,
there are no more obvious peaks
(say, above a matched-filter SNR $\rho_{\rm L1} > 8$)
other than GW230529 itself.

We would like to stress again that this simplified analysis
is \emph{not} how a full search should be done,
due to the absence of a statistically robust background estimation.
The final verdict regarding any possible lensed counterparts of GW230529
would have to wait for the release and analysis of the complete O4a data.
For now, preliminarily, we can say that we did not find any obvious GW signal
with a similar waveform to GW230529 within the 4096\,s of public data,
with a relative magnification $\mu_{\rm rel} \gtrsim \left(8/11.4\right)^2 \approx 1/2$
with respect to the known event.

\subsection{Signals with Negative Parity}\label{subsec:type_II}

In order to tell a type-II (negative parity) signal and an unlensed signal apart,
one will need to observe higher multipoles beyond the dominant quadrupolar GW emission
from the data \citep{Ezquiaga:2020gdt, Wang:2021kzt, Janquart:2021nus,Vijaykumar:2022dlp}.
Unfortunately, GW230529 is not particularly loud,
and therefore we do not expect to be able to do so a-priori.
Still, we went ahead and computed explicitly the Bayes factor
$\mathcal{B}^{\rm type\,II}_{\rm  unlensed}$
as the ratio of the evidence from a Bayesian parameter estimation
using type-II and normal unlensed GW signal waveforms, respectively,
as an indicator of the presence of imprints of a type-II signal in the data.
All analyses presented in this section were performed 
using the IMRPhenomXPHM waveform model~\citep{Pratten:2020ceb}.

As expected, the log Bayes factor was found to be
$\log_{10} \mathcal{B}^{\rm type\,II}_{\rm unlensed} = 0.04 \pm 0.10$ with the \texttt{hanabi.inference} code \citep{Lo:2021nae},
which is smaller than its uncertainty coming from the nested sampling algorithm
\citep{Skilling:2006gxv,Speagle:2019ivv}
when performing the Bayesian parameter estimation analyses.
Therefore, we cannot tell whether GW230529 is a lensed type-II GW signal or not.

\begin{figure}
\centering
\includegraphics[width=\columnwidth]{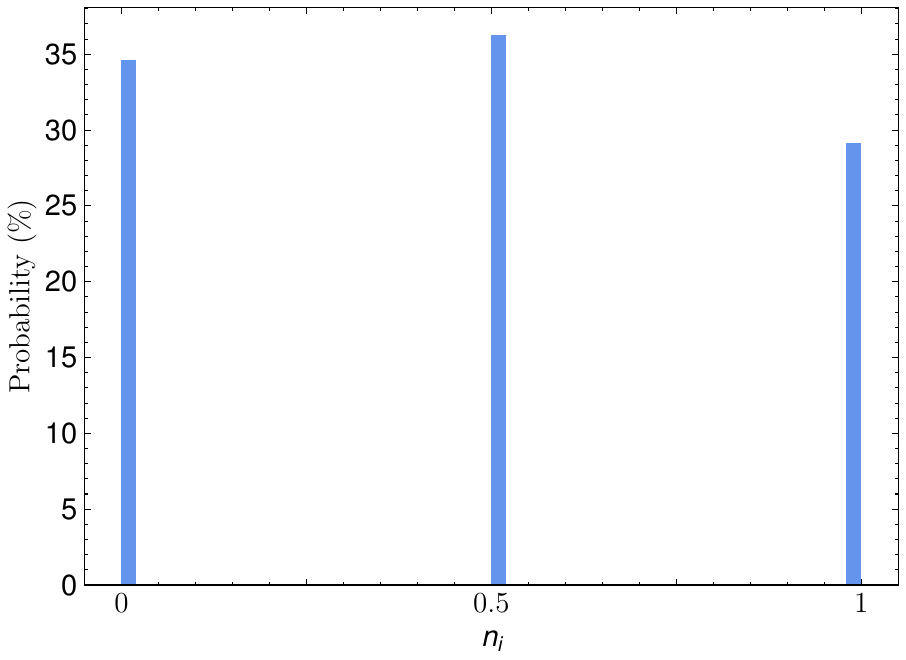}
\caption{\label{fig:golum_typeII} Posterior results from the \textsc{GOLUM} 
analysis for different image types.
The posterior is inconclusive, showing no preference for any hypothesis,
and thus we cannot tell whether GW230529 is a lensed type-II GW signal or not.}
\end{figure}

A similar analysis was performed using \textsc{GOLUM}~\citep{Janquart:2021qov, Janquart:2023osz} 
to compute the probabilities comparing the hypotheses of a lensed type-II signal against 
lensed type-I or type-III signals, which allows for comparing the signal under all 
lensing image type hypotheses.
The resulting posteriors for the Morse factor are shown
in Fig.~\ref{fig:golum_typeII}.
The results are consistent with the type-II vs. unlensed analysis, yielding a 
very inconclusive posterior with equal support for any hypothesis.
The resulting probabilities are $\ln(p(\mathrm{II vs. I})) = 0.05$ and $\ln(p(\mathrm{II vs. III})) = 0.22$.
This leads to the same conclusion that we cannot tell whether GW230529 is a lensed type-II GW signal or not.

\subsection{Microlensing Investigation}\label{subsec:microlensing}

To further investigate the possibility of GW230529 being a lensed signal,
we have also examined the data for any frequency-dependent waveform deformations
that could be signatures of microlensing. This examination was conducted
using the \textsc{Gravelamps} framework~\citep{Wright:2021cbn} for each of the
BBH, NSBH, and BNS cases. These were represented by the
\textsc{IMRPhenomXPHM}~\citep{Pratten:2020ceb},
\textsc{IMRPhenomNSBH}~\citep{Thompson:2020nei}, and
\textsc{IMRPhenomPv2\_NRTidalv2}~\citep{Dietrich:2019kaq} waveform approximants,
respectively. For this initial investigation, the isolated point-mass
microlensing model was used, mirroring the investigations conducted both more
widely~\citep{LIGOScientific:2023bwz} and in specific lensing investigations of
previous events~\citep{Janquart:2023mvf}.

None of these investigations yielded significant support for the microlensing
hypothesis with $\log_{10} \mathcal{B}^{\textrm{PM}}_{\textrm{U}}$ values of
$-0.014$, $0.098$, and $-0.019$ respectively. These values are all within the
expected error on these log Bayes factors from the nested sampling algorithm
\citep{Skilling:2006gxv,Speagle:2019ivv}.

Examination of the lens parameter posteriors for each of these investigations
revealed that in all three cases they reflect the priors, which is the
expectation for unlensed events. Similarly, comparisons of the chirp mass, mass
ratio, and luminosity distance posteriors---those parameters likely to be most
affected by the presence of microlensing---between the unlensed and lensed
investigations did not reveal any significant differences between the posteriors
consistent with an unlensed event. 

Therefore, we conclude there is no significant evidence that GW230529
displays any signatures of microlensing.

%% file: conclusions.tex
As already discussed in \citet{LIGOScientific:2024elc},
it is difficult to confirm the exact nature of
the component objects of the source of the GW230529 signal,
particularly of the heavier component
that appears to be within the purported 3--5\,$M_\odot$ mass gap.
In this work, we have performed various explorations 
into the gravitational lensing hypothesis
that would have made intrinsically lighter objects
appear heavier in the GW observation.
Overall, we find the lensing scenario is unlikely,
based on current astrophysical assumptions.

In particular, our simulation-based estimates of lensed detection rates
show that the chances for lensed BNS signals at O4a sensitivity
are very low, with relative rates between lensed and unlensed events
of $2\times 10^{-3}$ at most, with some variation depending 
on the assumed merger rate density. Hence, we would require numerous
GW230529-like events before there is a reasonable chance of
one being a lensed BNS. Moreover, by computing the Bayes factor accounting 
for the latest population models, we find a relatively strong disfavoring
of the lensing hypothesis. Additionally, we have studied what the lensing and
intrinsic source parameters would be under the lensing scenario.
Here we find the event would likely 
require high magnifications, sometimes beyond expectations, and 
have a substantial probability for the lighter component being a sub-solar mass object.
Therefore, adding the lensing hypothesis to the scenario generally does not
lead to a less exotic source system. We have also looked for additional lensing 
signatures such as the presence of a second image in the 4096\,s of publicly 
released data, the possibility of GW230529 being a type-II signal, and 
the presence of microlensing. In all cases, we find no evidence for 
these lensing-specific features. 

Based on these investigations, we conclude that GW230529
is unlikely a lensed BNS masquerading as a mass-gap NSBH merger.

%% file: acknowledgements.tex
We thank the members of the LIGO--Virgo--KAGRA lensing group for useful discussions.

This research has made use of data or software obtained from the Gravitational Wave Open Science Center (gwosc.org), a service of the LIGO Scientific Collaboration, the Virgo Collaboration, and KAGRA. This material is based upon work supported by NSF's LIGO Laboratory which is a major facility fully funded by the National Science Foundation, as well as the Science and Technology Facilities Council (STFC) of the United Kingdom, the Max-Planck-Society (MPS), and the State of Niedersachsen/Germany for support of the construction of Advanced LIGO and construction and operation of the GEO600 detector. Additional support for Advanced LIGO was provided by the Australian Research Council. Virgo is funded, through the European Gravitational Observatory (EGO), by the French Centre National de Recherche Scientifique (CNRS), the Italian Istituto Nazionale di Fisica Nucleare (INFN) and the Dutch Nikhef, with contributions by institutions from Belgium, Germany, Greece, Hungary, Ireland, Japan, Monaco, Poland, Portugal, Spain. KAGRA is supported by Ministry of Education, Culture, Sports, Science and Technology (MEXT), Japan Society for the Promotion of Science (JSPS) in Japan; National Research Foundation (NRF) and Ministry of Science and ICT (MSIT) in Korea; Academia Sinica (AS) and National Science and Technology Council (NSTC) in Taiwan.
The authors are grateful for computational resources provided by the
LIGO Laboratory
and supported by
National Science Foundation Grants PHY-0757058 and PHY-0823459.
We acknowledge the use of IUCAA LDG cluster Sarathi for the computational/numerical work.

D.\,Keitel, N.\,Singh and A.\,Heffernan were supported by the Universitat de les Illes Balears (UIB);
the Spanish Agencia Estatal de Investigaci{\'o}n grants
CNS2022-135440,
PID2022-138626NB-I00,
RED2022-134204-E,
RED2022-134411-T,
funded by MICIU/AEI/10.13039/501100011033,
the European Union NextGenerationEU/PRTR,
and the ERDF/EU;
and the Comunitat Aut{\`o}noma de les Illes Balears
through the Direcci{\'o} General de Recerca, Innovaci{\'o} I Transformaci{\'o} Digital
with funds from the Tourist Stay Tax Law
(PDR2020/11 - ITS2017-006)
as well as through the Conselleria d'Economia, Hisenda i Innovaci{\'o}
with grant numbers
SINCO2022/6719
(European Union NextGenerationEU/PRTR-C17.I1)
and SINCO2022/18146
(co-financed by the European Union
and FEDER Operational Program 2021-2027 of the Balearic Islands).

R.\,Lo, J.\,Chan and J.\,M.\, Ezquiaga were supported by the research grant no.~VIL37766 and no.~VIL53101 from Villum Fonden, the DNRF Chair program grant no. DNRF162 by the Danish National Research Foundation and the European Union's Horizon 2020 research and innovation programme under the Marie Sklodowska-Curie grant agreement No 101131233.
J.\,M.\, Ezquiaga is also supported by the Marie Sklodowska-Curie grant agreement No.~847523 INTERACTIONS.
The Tycho supercomputer hosted at the SCIENCE HPC center at the University of Copenhagen was used for supporting this work.

L. Uronen is supported by the Hong Kong PhD Fellowship Scheme (HKPFS) 
from the Hong Kong Research Grants Council (RGC). 
H. Phurailatpam, L. Uronen, and O.A. Hannuksela acknowledge support by grants from the RGC of Hong Kong 
(Project No. CUHK 14304622 and 14307923), 
the start-up grant from the Chinese University of Hong Kong, 
and the Direct Grant for Research from the Research Committee of The Chinese University of Hong Kong. 

M.W. is supported by the Science and Technology Facilities Council (STFC) of the United Kingdom Doctoral Training Grant ST/W507477/1.

T.\,Bulik was supported by the National Science Centre, Poland through grants 2023/49/B/ST9/02777 and 2018/30/A/ST9/00050.

This paper has been assigned document number LIGO-\dcc.